\documentclass[]{aa}

\usepackage{amssymb,amsmath}
\usepackage{subfigure,multirow}
\usepackage{graphicx}
\usepackage{natbib}
\usepackage{xspace}
\usepackage{txfonts}

\newcommand{\fsky}{\ensuremath{f_\text{sky}}}
\newcommand{\smica}{\textsc{Smica}}
\newcommand{\Bmode}{B-mode}
\newcommand{\Bmodes}{B-modes}

\newcommand{\Emodes}{E-modes}
\newcommand{\modcl}{C_\ell}
\newcommand{\estcl}{\hat{C}_\ell}
\newcommand{\shapecl}{\mathcal{S}_\ell}
\newcommand{\modnl}{\mathcal{N}_\ell}
\newcommand{\Acmb}{\ensuremath{\vec{A}_\text{cmb}}}
\newcommand{\Rcont}{\ensuremath{\tens{N}_\ell}}

\newcommand{\loglik}{\ln \mathcal{L}}
\newcommand{\lmin}{\ell_{\min}}
\newcommand{\lmax}{\ell_{\max}} 
\newcommand{\spos}{\xi}
\newcommand{\lm}{\ensuremath{{\ell m}}} 
\makeatletter
\newcommand*{\R}{\@ifstar\Rest\Rnorm}
\newcommand*{\Rnorm}[1][]{\ensuremath{\tens{R}^\text{#1}}}
\newcommand*{\Rest}[1][]{\ensuremath{\widehat{\tens{R}}^\text{#1}}}
\makeatother
\DeclareMathOperator{\expectation}{\mathbf{E}}
\DeclareMathOperator{\trace}{trace}
\DeclareMathOperator{\diag}{diag}
\DeclareMathOperator{\logdet}{\log\,\det}
\DeclareMathOperator{\var}{var}
\newcommand{\expect}[1]{\ensuremath{\expectation \left( #1 \right) }}
\newcommand{\KL}[2]{\ensuremath{K \left( #1, #2 \right) }}
\newcommand{\domain}{\ensuremath{\mathcal{D}_q}}
\providecommand{\micron}{\ensuremath{\mathrm{\mu m} }}
\newcommand{\microK}{\ensuremath{\mathrm{\mu K} }}

\begin{document}
\title{Measuring the tensor to scalar ratio from CMB B-modes in presence of foregrounds}

\author{Marc Betoule\inst{1}
\and Elena Pierpaoli\inst{2}
\and Jacques Delabrouille\inst{1}
\and Maude Le Jeune\inst{1}
\and Jean-Fran\c cois Cardoso\inst{1}
 }
\institute{AstroParticule et Cosmologie (APC), CNRS : UMR 7164 - Universit\'e Denis Diderot - Paris 7 - Observatoire de Paris, FRANCE
\and University of Southern California, Los Angeles, CA, 90089-0484, USA}

\abstract{}
{We investigate the impact of polarised foreground emission on the
  performances of future CMB experiments aiming the detection of
  primordial tensor fluctuations in the early universe. In particular,
  we study the accuracy that can be achieved in measuring the tensor--to--scalar
  ratio $r$ in presence of
  foregrounds.}
{We design a component separation pipeline, based on the \smica\
  method, aimed at estimating $r$ and the foreground contamination
  from the data with no prior assumption on the frequency dependence
  or spatial distribution of the foregrounds.  We derive error bars
  accounting for the uncertainty on foreground contribution. We use
  the current knowledge of galactic and extra-galactic foregrounds as
  implemented in the Planck Sky Model (PSM), to build simulations of
  the sky emission.  We apply the method to simulated observations of
  this modelled sky emission, for various experimental setups.}
{Our method, with Planck data, permits us to detect $r=0.1$ from
  \Bmodes\ only at more than 3$\sigma$.  With a future dedicated space
  experiment, as EPIC, we can measure $r=0.001$ at $\sim 6 \sigma$ for
  the most ambitious mission designs.  Most of the sensitivity to $r$
  comes from scales $20 \le \ell \le 150$ for high $r$ values,
  shifting to lower $\ell$'s for progressively smaller $r$. This shows
  that large scale foreground emission doesn't prevent a proper
  measurement of the reionisation bump for full sky experiment.  We
  also investigate the observation of a small but clean part of the
  sky.  We show that diffuse foregrounds remain a concern for a
  sensitive ground--based experiment with a limited frequency coverage
  when measuring $r < 0.1$. Using the Planck data as additional
  frequency channels to constrain the foregrounds in such
  ground--based observations reduces the error by a factor two but
  does not allow to detect $r=0.01$.  An alternate strategy, based on
  a deep field space mission with a wide frequency coverage, would
  allow us to deal with diffuse foregrounds efficiently, but is in
  return quite sensitive to lensing contamination.  In the contrary,
  we show that all-sky missions are nearly insensitive to small scale
  contamination (point sources and lensing) if the statistical
  contribution of such foregrounds can be modelled accurately.  Our
  results do not significantly depend on the overall level and
  frequency dependence of the diffused foreground model, when varied
  within the limits allowed by current observations.  }
{}


\titlerunning{$T/S$ measurements in presence of foregrounds}
\authorrunning{Betoule et al.}

\maketitle

\section{Introduction}

After the success of the WMAP space mission in mapping the Cosmic
Microwave Background (CMB) temperature anisotropies, much attention
now turns towards the challenge of measuring CMB polarisation, in
particular pseudo-scalar polarisation modes (the \Bmodes) of
primordial origin. These \Bmodes\ offer one of the best options to
constrain inflationary models
\citep{1997PhRvL..78.2054S,1997NewA....2..323H,1997PhRvL..78.2058K,1998PhRvD..57..685K,2008arXiv0810.3022B}.

First polarisation measurements have already been obtained by a number
of instruments
\citep{2002Natur.420..772K,2005AAS...20710007S,2007ApJS..170..335P},
but no detection of \Bmodes\ has been claimed yet.  While several
ground--based and balloon--borne experiments are already operational,
or in construction, no CMB--dedicated space-mission is planned after
Planck at the present time: whether there should be one for CMB
\Bmodes, and how it should be designed, are still open questions.
 
As CMB polarisation anisotropies are expected to be significantly
smaller than temperature anisotropies (a few per cent at most),
improving detector sensitivities is the first major challenge towards
measuring CMB polarisation B-modes. It is not, however, the only
one. Foreground emissions from the galactic interstellar medium (ISM)
and from extra-galactic objects (galaxies and clusters of galaxies)
superimpose to the CMB. Most foregrounds are expected to emit
polarised light, with a polarisation fraction typically comparable, or
larger, than that of the CMB. Component separation (disentangling CMB
emission from all these foregrounds) is needed to extract cosmological
information from observed frequency maps.  The situation is
particularly severe for the \Bmodes\ of CMB polarisation, which will
be, if measurable, sub-dominant at every scale and every frequency.
  
\begin{figure*}
  \includegraphics{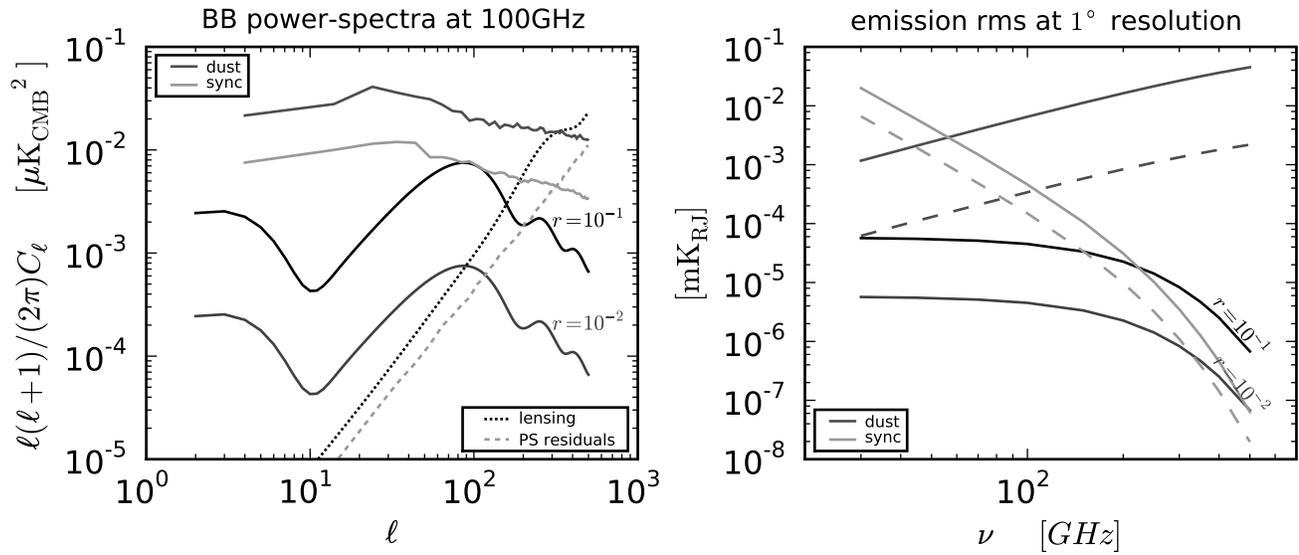}
  \caption{Respective emission levels of the various components as predicted by the PSM.
 Left: predicted power spectra of the various components at 100~GHz, compared to CMB and lensing level for standard cosmology and various values of $r$ ($\tau = 0.07$, and other cosmological parameters follow \cite{2008arXiv0803.0586D}). The power spectra of diffuse galactic foregrounds are computed using the cleanest 55\% of the polarised sky. The power spectrum from residual point sources is computed assuming that all sources brighter than 500~mJy (in temperature) in one of the Planck channels have been cut out.
Right: typical frequency-dependence of the contributions to B-type polarisation of CMB, synchrotron and dust, at 1 degree resolution. The dashed lines correspond to the mean level of fluctuation as computed outside the mask used for the power spectra shown in the right panel.}
  \label{fig:psm}
\end{figure*}

The main objective of this paper is to evaluate the accuracy with
which various upcoming or planned experiments can measure $r$ in
presence of foregrounds.  This problem has been addressed before:
\citet{2005MNRAS.360..935T} investigate the lower bound for $r$ that
can be achieved considering a simple foreground cleaning technique,
based on the extrapolation of foreground templates and subtraction
from a channel dedicated to CMB measurement;
\citet{2006JCAP...01..019V} assume foreground residuals at a known
level in a cleaned map, treat them as additional Gaussian noise, and
compute the error on $r$ due to such excess noise;
\citet{2007PhRvD..75h3508A} investigate how best to select the
frequency bands of an instrument, and how to distribute a fixed number
of detectors among them, to maximally reject galactic foreground
contamination. This latter analysis is based on an Internal Linear
Combination cleaning technique similar to the one of
\citet{2003PhRvD..68l3523T} on WMAP temperature anisotropy data.  The
two last studies assume somehow that the residual contamination level
is perfectly known -- an information which is used to derive error
bars on $r$.

In this paper, we relax this assumption and propose a method to
estimate the uncertainty on residual contamination from the data
themselves, as would be the case for real data analysis.  We test our
method on semi-realistic simulated data sets, including CMB and
realistic foreground emission, as well as simple instrumental noise.
We study a variety of experimental designs and foreground mixtures.

This paper is organised as follows: the next section
(Sect. \ref{sec:fg}) deals with polarised foregrounds and presents the
galactic emission model used in this work. In section
\ref{sec:method}, we propose a method, using the most recent version
of the \smica\ component separation framework
\citep{2008arXiv0803.1814C}, to provide measurements of the tensor to
scalar ratio in presence of foregrounds. In section \ref{sec:results},
we present the results obtained by applying the method to various
experimental designs. Section \ref{sec:discussion} discusses the
reliability of the method (and of our conclusions) against various
issues, in particular modelling uncertainty. Main results are
summarised in section \ref{sec:conclusion}.

\section{Modelling polarised sky emission}
\label{sec:fg}

Several processes contribute to the total sky emission in the
frequency range of interest for CMB observation (typically between 30
and 300~GHz). Foreground emission arises from the galactic
interstellar medium (ISM), from extra-galactic objects, and from
distortions of the CMB itself through its interaction with structures
in the nearby universe.  Although the physical processes involved and
the general emission mechanisms are mostly understood, specifics of
these polarised emissions in the millimetre range remain poorly known
as few actual observations, on a significant enough part of the sky,
have been made.

Diffuse emission from the ISM arises through synchrotron emission from
energetic electrons, through free--free emission, and through
grey-body emission of a population of dust grains. Small spinning dust
grains with a dipole electric moment may also emit significantly in
the radio domain \citep{1998ApJ...508..157D}. Among those processes,
dust and synchrotron emissions are thought to be significantly
polarised. Galactic emission also includes contributions from compact
regions such as supernovae remnants and molecular clouds, which have
specific emission properties.

Extra-galactic objects emit via a number of different mechanisms, each
of them having its own spectral energy distribution and polarisation
properties.

Finally, the CMB polarisation spectra are modified by the interactions
of the CMB photons on their way from the last scattering
surface. Reionisation, in particular, re-injects power in polarisation
on large scales by late-time scattering of CMB photons.  This produces
a distinctive feature, the reionisation bump, in the CMB \Bmode\
spectrum at low $\ell$. Other interactions with the latter universe,
and in particular lensing, contribute to hinder the measurement of the
primordial signal. The lensing effect is particularly important on
smaller scales as it converts a part of the dominant E-mode power into
B-mode.

In the following, we review the identified polarisation processes and
detail the model used for the present work, with a special emphasis on
\Bmodes. We also discuss main sources of uncertainty in the model, as
a basis for evaluating their impact on the conclusions of this paper.

Our simulations are based on the Planck Sky Model (PSM), a sky
emission simulation tool developed by the Planck collaboration for
pre-launch preparation of Planck data analysis
\citep{Delabrouille09}. Figure \ref{fig:psm} gives an overview of
foregrounds as included in our baseline model. Diffuse galactic emission from synchrotron and
dust dominates at all frequencies and all scales, with a minimum (relative to CMB) between 60 and 80 GHz, depending on the galactic cut. Contaminations by lensing and a point source background are lower than primordial CMB for $r > 0.01$ and for $\ell < 100$, but should clearly be taken into account in attempts to measure $r < 0.01$.

\subsection{Synchrotron}
\label{sec:syncphys}

Cosmic ray electrons spiralling in the galactic magnetic field produce
highly polarised synchrotron emission
(e.g. \citet{1979rpa..book.....R}). This is the dominant contaminant
of the polarised CMB signal at low frequency (\(\lesssim
80\,\text{GHz}\)), as can be seen in the right panel of Fig. \ref{fig:psm}. In the
frequency range of interest for CMB observations, measurements of this
emission have been provided, both in temperature and polarisation, by
WMAP \citep{2007ApJS..170..335P,2008arXiv0803.0715G}. The intensity of
the synchrotron emission depends on the cosmic ray density
\(n_e\), and on the strength of the magnetic field perpendicularly to the
line of sight. Its frequency scaling and its intrinsic polarisation
fraction \(f_s\) depend on the energy distribution of the cosmic
rays. 

\subsubsection{Synchrotron emission law} 

For electron density following a power law of index \(p\),
\(n_e(E) \varpropto E^{-p} \), the synchrotron frequency dependence is
also a power law, of index \(\beta_s = -(p + 3)/2\):
\begin{equation}
S(\nu) = S(\nu_0) (\nu/\nu_0)^{\beta_s}
\label{eq:syncpowlaw}
\end{equation}
where the spectral index, \(\beta_s\), is equal to $-3$ for a typical value \(p = 3\).

The synchrotron spectral index depends significantly on
cosmic ray properties. It varies with the direction of the sky, and
possibly, with the frequency of observation (see
e.g. \citet{2007ARNPS..57..285S} for a review of propagation and
interaction processes of cosmic rays in the galaxy).  

For a multi-channel experiment, the consequence of this is a
decrease of the coherence of the synchrotron emission across channels, i.e. the
correlation between the synchrotron emission in the various frequency
bands of observation will be below unity.

Observational constraints have been put on the synchrotron emission law.  
A template of synchrotron emission intensity at 408~MHz has been
provided by \citet{1982A&AS...47....1H}. Combining this map with sky
surveys at 1.4~GHz \citep{1986A&AS...63..205R} and 2.3~GHz
\citep{1998MNRAS.297..977J}, \citet{2002A&A...387...82G} and
\citet{2003A&A...410..847P} have derived nearly full sky spectral
index maps.  Using the measurement from WMAP,
\citet{2003ApJS..148...97B} derived the spectral index between 408~MHz
and 23~GHz. Compared to the former results, it showed a
significant steepening toward \(\beta_s = -3\) around 20~GHz, and a
strong galactic plane feature with flatter spectral index. This
feature was first interpreted as a flatter cosmic ray distribution in
star forming regions.  Recently, however, taking into account the presence, at
23 GHz, of additional contribution from a possible anomalous emission
correlated with the dust column density, \citet{2008arXiv0802.3345M}
found no such pronounced galactic feature, in better agreement with
lower frequency results. The spectral index map obtained in this way
is consistent with \(\beta_s = -3 \pm 0.06\).

There is, hence, still significant uncertainty on the exact variability of the synchrotron 
spectral index, and in the amplitude of the steepening if any. 

\subsubsection{Synchrotron polarisation} 

If the electron density follows a power law of index \(p\),
the synchrotron polarisation fraction reads:
\begin{equation}
f_s = 3(p + 1)/(3p + 7)
\label{eq:syncpolfrac}
\end{equation}

For \(p = 3\), we get \(f_s = 0.75\), a polarisation fraction which
varies slowly for small variations of \(p\). Consequently, the
intrinsic synchrotron polarisation fraction should be close to
constant on the sky. However, geometric depolarisation arises due to
variations of the polarisation angle along the line of sight, partial
cancellation of polarisation occurring for superposition of emission
with orthogonal polarisation directions. Current measurements show
variations of the observed polarisation value from about 10\% near the
galactic plane, to 30-50 \% at intermediate to high galactic latitudes
\citep{Macellari08}.

\subsubsection{Our model of synchrotron} 

In summary, the \Bmode\ intensity of the synchrotron emission is
modulated by the density of cosmic rays, the slope of their spectra,
the intensity of the magnetic field, its orientation, and the
coherence of the orientation along the line of sight. This makes the amplitude and
frequency scaling of the polarised synchrotron signal dependant on the sky
position in a rather complex way.

For the purpose of the present work, we mostly follow
\cite{2008arXiv0802.3345M} model 4, using the same synchrotron
spectral index map, and the synchrotron polarised template at 23~GHz
measured by WMAP. This allows the definition of a pixel-dependent
geometric depolarisation factor $g(\xi)$, computed as the ratio
between the polarisation expected theoretically from
Eq. \ref{eq:syncpolfrac}, and the polarisation actually observed. This
depolarisation, assumed to be due to varying orientations of the
galactic magnetic field along the line of sight, is used also for
modelling polarised dust emission (see below).

As an additional refinement, we also investigate the impact of a slightly modified frequency dependence with a running spectral index in Sect. \ref{sec:discussion}. For this purpose, the synchrotron emission Stokes parameters ($S_{\nu}^X(\spos)$ for $X \in \lbrace Q,U\rbrace$), at frequency $\nu$ and in direction $\spos$ on the sky, will be modelled instead as:
\begin{equation}
\label{eq:syncelaw}
S_{\nu}^X(\spos) = S^X_{\nu_0}(\spos) \left(\cfrac{\nu}{\nu_0}\right)^{\beta_s(\spos)+C(\spos) \log(\nu / \nu_1)}
\end{equation}
where $S^X_{\nu_0}(\spos)$ is the WMAP measurement at $\nu_0 = 23 \text{GHz}$, $\beta_s$ the synchrotron spectral index map \citep{2008arXiv0802.3345M}, and $C(\spos)$ a synthetic template of the curvature of the synchrotron spectral index.

The reconstructed \Bmodes\ map of the synchrotron-dominated sky emission at 30~GHz is shown in Fig. \ref{fig:polarizedfg}.

\subsection{Dust}
\label{sec:dustphys}

The thermal emission from heated dust grains is the dominant galactic
signal at frequencies higher than 100~GHz
(Fig. \ref{fig:psm}). Polarisation of starlight by dust grains
indicates partial alignment of elongated grains with the galactic
magnetic field (see \citet{2007JQSRT.106..225L} for a review of
possible alignment mechanisms). Partial alignment of grains should
also result in polarisation of the far infrared dust emission.

Contributions from a wide range of grain sizes and compositions are
required to explain the infrared spectrum of dust emission from 3 to
1000~\(\micron\) \citep{1990A&A...237..215D,2001ApJ...554..778L}.  At
long wavelengths of interest for CMB
observations (above 100~\(\micron\)), the emission from big grains, at equilibrium with the
interstellar radiation field, should
dominate. 

\subsubsection{Dust thermal emission law} 

There is no single theoretical emission law for dust, which is
composed of many different populations of particles of matter. On
average, an emission law can be fit to observational data. In the
frequency range of interest for CMB observations,
\citet{1999ApJ...524..867F} have shown that the dust emission in
intensity is well modelled by emission from a two components mixture
of silicate and carbon grains. For both components, the thermal
emission spectrum is modelled as a modified grey-body emission, \(
D_\nu \sim B_\nu(T) v^\alpha \), with different emissivity spectral
index \(\alpha\) and different equilibrium temperature $T$.
 
\subsubsection{Dust polarisation} 

So far, dust polarisation measurements have been mostly concentrated on specific regions of emission, with the exception of the Archeops balloon-borne experiment \citep{2004A&A...424..571B}, which has mapped the emission at 353~GHz on a significant part of the sky, showing a polarisation fraction around 4-5\% and up to 10\% in some clouds. This is in rough agreement with what could be expected from polarisation of starlight \citep{2002ApJ...564..762F,2008arXiv0809.2094D}.
\cite{Macellari08} show that dust fractional polarisation in WMAP5 data depends on both frequency 
and latitude, but is typically about 3\% and anyway below 7\%. 

\citet{2008arXiv0809.2094D} have shown that for particular mixtures of dust grains, the intrinsic polarisation of the dust emission could vary significantly with frequency in the 100-800~GHz range. Geometrical depolarisation caused by integration along the line of sight also lowers the observed polarisation fraction.

\subsubsection{Our model of dust} 

To summarise, dust produces polarised light depending on grains shape, size, composition, temperature and environment. The polarised light is then observed after integration along a line of sight. Hence, the observed polarisation fraction of dust depends on its three-dimensional distribution, and of the geometry of the galactic magnetic field. This produces a complex pattern which is likely to be only partially coherent from one channel to another. 

\begin{figure}

  \includegraphics[angle=90]{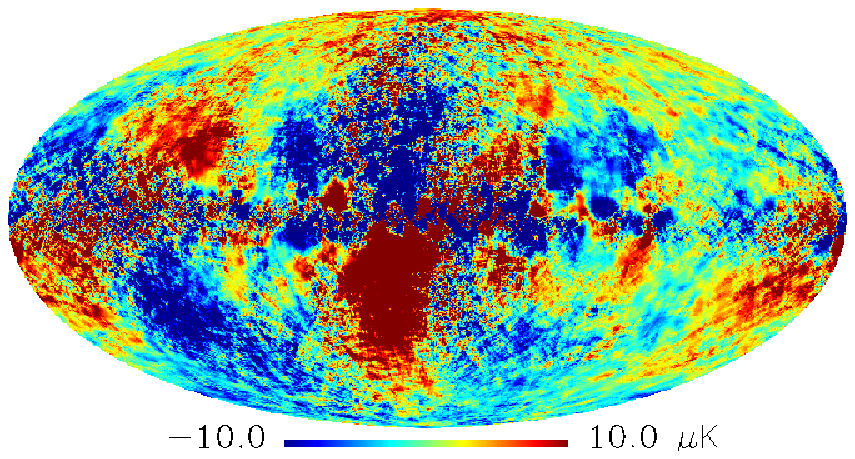}

  \includegraphics[angle=90]{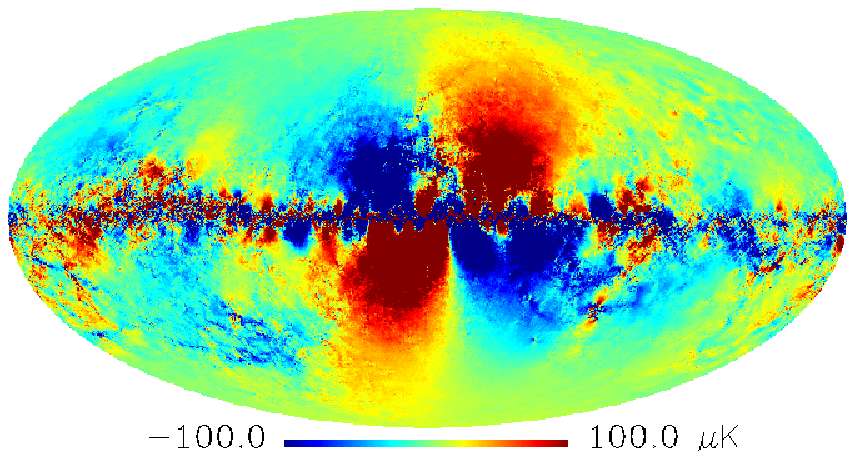}
  \caption{\Bmodes\ of the galactic foreground maps (synchrotron + dust) as simulated using v1.6.4 of the PSM. Top: synchrotron-dominated emission at 30~GHz, Bottom: dust-dominated emission at 340~GHz. In spite of the fact that the direction of polarisation of both processes is determined by the same galactic magnetic field, differences in the 3-D distributions and in the depolarisation factors result in quite different \Bmode\ polarisation patterns.}
  \label{fig:polarizedfg}
\end{figure}

Making use of the available data, the PSM models polarised thermal
dust emission by extrapolating dust intensity to polarisation
intensity assuming an intrinsic polarisation fraction $f_d$ constant
across frequencies. This value is set to $f_d = 0.12$ to be
consistent with maximum values observed by Archeops
\citep{2004A&A...424..571B} and is in good agreement with the WMAP
94~GHz measurement. The dust intensity ($D_\nu^T$), traced by the
template map at 100~\(\micron\) from \citet{1998ApJ...500..525S}, is
extrapolated using \citet[model \#7]{1999ApJ...524..867F} to
frequencies of interest.  The stokes \(Q\) and \(U\) parameters
(respectively \(D^Q\) and \(D^U\)) are then obtained as:
\begin{eqnarray}
  D^Q_\nu(\spos) &= f_d \, g(\spos) \, D^T_\nu(\spos) \, \cos(2\gamma(\spos))\\
  D^U_\nu(\spos) &= f_d \, g(\spos) \, D^T_\nu(\spos) \, \sin(2\gamma(\spos))
\end{eqnarray}
The geometric `depolarisation' factor $g$ is a modified version of the
synchrotron depolarisation factor (computed from WMAP measurements).
Modifications account for differences of spatial distribution between
dust grains and energetic electrons, and are computed using the magnetic
field model presented in \cite{2008arXiv0802.3345M}.  
The polarisation angle \(\gamma\) is obtained from the magnetic field
model on large scales and from synchrotron measurements in WMAP on
scales smaller than 5 degrees.  Figure \ref{fig:polarizedfg} shows the
\Bmodes\ of dust at 340~GHz using this model.

\subsubsection{Anomalous dust} 

If the anomalous dust emission, which may account for a significant
part of the intensity emission in the range 10-30~GHz
\citep{2004ApJ...614..186F,2004ApJ...606L..89D,2008arXiv0802.3345M},
can be interpreted as spinning dust grains emission
\citep{1998ApJ...508..157D}, it should be slightly polarised under
35~GHz \citep{2006ApJ...645L.141B}, and only marginally polarised at
higher frequencies \citep{2003NewAR..47.1107L}.  For this reason, it
is neglected (and not modelled) here.  However, we should keep in mind
that there exist other possible emission processes for dust, like the
magneto-dipole mechanism, which can produce highly polarised
radiation, and could thus contribute significantly to dust
polarisation at low frequencies, even if sub-dominant in intensity
\citep{2003NewAR..47.1107L}.

\subsection{Other processes}
\label{sec:other}
The left panel in Fig. \ref{fig:psm} presents the respective
contribution from the various foregrounds as predicted by the PSM at
100~GHz.  Synchrotron and dust polarised emission, being by far the
strongest contaminants on large scales, are expected to be the main
foregrounds for the measurement of primordial \Bmodes.  In this work,
we thus mainly focus on the separation from these two diffuse
contaminants.  However, other processes yielding polarised signals at
levels comparable with either the signal of interest, or with the
sensitivity of the instrument used for \Bmode\ observation, have to be
taken into account.

\subsubsection{Free-free}

Free-free emission is assumed unpolarised to first order (the emission
process is not intrinsically linearly polarised), even if, in
principle, low level polarisation by Compton scattering could exist at
the edge of dense ionised regions.  In WMAP data analysis,
\cite{Macellari08} find an upper limit of 1\% for free--free
polarisation. At this level, free-free would have to be taken into
account for measuring CMB \Bmodes\ for low values of $r$. As this is
just an upper limit however, no polarised free-free is considered for
the present work.

\subsubsection{Extra-galactic sources}

Polarised emission from extra-galactic sources is expected to be faint
below the degree scale. \citet{2005MNRAS.360..935T}, however, estimate
that radio sources become the major contaminant after subtraction of
the galactic foregrounds. It is, hence, an important foreground at
high galactic latitudes.  In addition, the point source contribution
involves a wide range of emission processes and superposition of
emissions from several sources, which makes this foreground poorly
coherent across frequencies, and hence difficult to subtract using
methods relying on the extrapolation of template emission maps.

The Planck Sky Model provides estimates of the point source polarised
emission.  Source counts are in agreement with the prediction of
\citet{2005A&A...431..893D}, and with WMAP data.  For radio-sources,
the degree of polarisation for each source is randomly drawn from the
observed distribution at 20~GHz \citep{2004A&A...415..549R}.  For
infrared sources, a distribution with mean polarisation degree of 0.01
is assumed.  For both populations, polarisation angles are uniformly
drawn in $[0-2\pi[$.  The emission of a number of known galactic point
sources is also included in PSM simulations.

\subsubsection{Lensing}

The last main contaminant to the primordial \Bmode\ signal is
lensing-induced B-type polarisation, the level of which should be of
the same order as that of point sources (left panel of
Fig. \ref{fig:psm}).  For the present work, no sophisticated lensing
cleaning method is used. Lensing effects are modelled and taken into
account only at the power spectrum level and computed using the CAMB
software package,\footnote{http://camb.info} based itself on the
CMBFAST software \citep{1998ApJ...494..491Z,2000ApJS..129..431Z}.

\subsubsection{Polarised Sunyaev-Zel'dovich effect}

The polarised Sunyaev Zel'dovich effect \citep{1999MNRAS.310..765S,1999MNRAS.305L..27A,Seto05}, is expected to be very sub-dominant and is neglected here.

\subsection{Uncertainties on the foreground model}
\label{sec:moderrors}
Due to the relative lack of experimental constraints from observation
at millimetre wavelengths, uncertainties on the foreground model are
large. The situation will not drastically improve before
the Planck mission provides new observations of polarised
foregrounds. It is thus very important to evaluate, at least
qualitatively, the impact of such uncertainties on component
separation errors for \Bmode\ measurements.

We may distinguish two types of uncertainties, which impact
differently the separation of CMB from foregrounds.  One concerns the
level of foreground emission, the other its complexity.

Quite reliable constraints on the emission level of polarised
synchrotron at 23~GHz are available with the WMAP measurement, up to
the few degrees scale. Extrapolation to other frequencies and smaller
angular scales may be somewhat insecure, but uncertainties take place
where this emission becomes weak and sub-dominant. The situation is
worse for the polarised dust emission, which is only weakly
constrained from WMAP and Archeops at 94 and 353~GHz. The overall
level of polarisation is constrained only in the galactic plane, and
its angular spectrum is only roughly estimated. In addition,
variations of the polarisation fraction \citep{2008arXiv0809.2094D}
may introduce significant deviations to the frequency scaling of dust
\Bmodes.

Several processes make the spectral indexes of dust and synchrotron
vary both in space and frequency. Some of this complexity is included
in our baseline model, but some aspects, like the dependence of the
dust polarisation fraction with frequency and the steepening of the
synchrotron spectral index, remain poorly known and are not modelled
in our main set of simulations.  In addition, uncharacterised emission
processes have been neglected. This is the case for anomalous dust, or
polarisation of the free-free emission through Compton scattering. If
such additional processes for polarised emission exist, even at a low
level, they would decrease the coherence of galactic foreground
emission between frequency channels, and hence our ability to predict
the emission in one channel knowing it in the others -- a point of
much importance for {\emph {any}} component separation method based on
the combination of multi-frequency observations.

The component separation as performed in this paper, hence, is
obviously sensitive to these hypotheses. We will dedicate a part of
the discussion to assess the impact of such modelling errors on our
conclusions.


\section{Estimating $r$ with contaminants}
\label{sec:method}

Let us now turn to a presentation of the component separation (and
parameter estimation) method used to derive forecasts on the tensor to
scalar ratio measurements.

Note that in principle, the best analysis of CMB observations should
simultaneously exploit measurements of all fields ($T$, $E$, and $B$),
as investigated already by \citet{2007MNRAS.376..739A}. Their work,
however, addresses an idealised problem. For component separation of
temperature and polarisation together, the best approach is likely to
depend on the detailed properties of the foregrounds (in particular on
any differences, even small, between foreground emissions laws in
temperature and in polarisation) and of the instrument (in particular
noise correlations, and instrumental systematics). None of this is
available for the present study. For this reason, we perform component
separation in \Bmode\ maps only. Additional issues such as
disentangling $E$ from $B$ in cases of partial sky coverage for
instance, or in presence of instrumental systematic effects, are not
investigated here either. Relevant work can be found in
\citet{2002AIPC..609..209K,2003NewAR..47..995C,2003PhRvD..67d3004H,2007A&A...464..405R}.

For low values of tensor fluctuations, the constraint on $r$ is
expected to come primarily from the \Bmode\ polarisation. \Bmodes\
indeed are not affected by the cosmic variance of the scalar
perturbations, contrarily to \Emodes\ and temperature anisotropies. In
return, \Bmode\ signal would be low and should bring little constraint
on cosmological parameters other than $r$ (and, possibly, the tensor
spectral index $n_t$, although this additional parameter is not
considered here). Decoupling the estimation of $r$ (from \Bmodes\
only) from the estimation of other cosmological parameters (from
temperature anisotropies, from \Emodes, and from additional
cosmological probes) thus becomes a reasonable hypothesis for small
values of $r$. As we are primarily interested in accurate handling of
the foreground emission, we will make the assumption that all
cosmological parameters but $r$ are perfectly known. Further
investigation of the coupling between cosmological parameters can be
found in \citet{colombo2008,2006JCAP...01..019V}, and this question is
discussed a bit further in Sect. \ref{sec:tau}.

\subsection{Simplified approaches}

\subsubsection{Single noisy map}
\label{sec:singlemap}

The first obstacle driving the performance of an experiment being the
instrumental noise, it is interesting to recall the limit on $r$
achievable in absence of foreground contamination in the observations.

We thus consider first a single frequency observation of the CMB, contaminated by a noise term $n$:
\begin{equation}
  \label{eq:modelsinglemap}
  x(\spos) =   x^\text{cmb}(\spos) +  n(\spos)
\end{equation}
where $\spos$ denotes the direction in the sky. Assuming that $n$ is uncorrelated with the CMB, the
power spectra of the map reads:
\[
\modcl = r \shapecl + \modnl  
\]
where $\shapecl $ is the shape of the CMB power-spectrum (as set by other cosmological
parameters), and $\modnl$ the power of the noise contamination.
Neglecting mode to mode mixing effects from a mask (if any), or in general from incomplete sky coverage, and assuming
that $n$ can be modelled as a Gaussian process, the log-likelihood
function for the measured angular power spectrum reads:
\begin{equation}\label{eq:loglikesingle}
  -2 \loglik
  =
  \sum_\ell (2\ell+1) \fsky 
  \left[ \ln\left(\frac{\modcl}{\estcl}\right)+\frac{\estcl}{\modcl}\right] 
  +\mathrm{const.}
\end{equation}

The smallest achievable variance $\sigma_r^2$ in estimating $r$ is the
inverse of the Fisher information $\mathcal{I} =
-\expect{\frac{\partial^2\loglik} {\partial r^2}}$ which takes the
form:
\begin{equation}\label{eq:roughsigma}
  \sigma_r^{-2} 
  = 
  \sum_{\ell=\lmin}^{\lmax} 
  \frac{2\ell +1}{2}
  \fsky 
  \left(\frac{\shapecl}{r \shapecl +\modnl} \right)^2
\end{equation}
For a detector (or a set of detectors at the same frequency) of noise
equivalent temperature \(s\) (in \(\mu K \sqrt{s}\)), and a mission
duration of \(t_s\) seconds, the detector noise power spectrum is \(
\modnl = \frac{4\pi s^2}{B_\ell^2 t_s} \, \microK^2 \), with
\(B_\ell\) denoting the beam transfer function of the detector.

A similar approach to estimating $ \sigma_r$ is used in \cite{2006JCAP...01..019V} where a
single `cleaned' map is considered. This map is obtained by optimal
combination of the detectors with respect to the noise and cleaned
from foregrounds up to a certain level of residuals, which are accounted for as an
extra Gaussian noise.

\subsubsection{Multi-map estimation}
\label{sec:multimap}

Alternatively, we may consider observations in $F$ frequency bands,
and form the $F \times 1$ vector of data \(\vec{x}(\spos) \), assuming
that each frequency is contaminated by $\vec{x}^\mathrm{cont}$.  This
term includes all contaminations (foregrounds, noise, etc...).  In the
harmonic domain, denoting $\vec{A}_\text{cmb}$ the emission law of the
CMB (the unit vector when working in thermodynamic units):
\begin{equation}
  \label{eq:modelnoisonly}
  \vec{a}_\lm  =   \vec{A}_\text{cmb} a_\lm^\mathrm{cmb} +  \vec{a}_\lm^\mathrm{cont} 
\end{equation}
We then consider the $F \times F$ spectral covariance matrix $\R_\ell$
containing auto and cross-spectra.  The CMB signal being uncorrelated
with the contaminants, one has:
\begin{equation}  \label{eq:modelspec}
  \R_\ell = \R[cmb]_\ell + \Rcont 
\end{equation}
with the CMB contribution modelled as
\begin{equation}
  \label{eq:modelcmb}
  \R[cmb]_\ell(r) = r \shapecl \vec{A}_\text{cmb} \vec{A}_\text{cmb}^\dag  
\end{equation}
and all contaminations contributing a term $\Rcont$ to be
discussed later.  The dagger ($\dag$) denotes the conjugate transpose for complex vectors and matrices, and the transpose for real matrices (as $\vec{A}_\text{cmb}$).

In the approximation that contaminants are Gaussian (and, here, stationary) but correlated, all the relevant information about the CMB is preserved by combining
all the channels into a single filtered map. In the harmonic domain,
the filtering operation reads:
\begin{displaymath}
  \tilde a_\lm 
  = \vec{W_\ell} \vec{a}_\lm
  = a_\lm^\mathrm{cmb} +  \vec{W_\ell} \vec{a}_\lm^\mathrm{cont}
\end{displaymath}
with
\begin{equation}
  \vec{W}_\ell 
  = \frac
  {\Acmb^\dag {\Rcont}^{-1}}
  {\Acmb^\dag {\Rcont}^{-1}\Acmb}
  \label{eq:well-ideal}
\end{equation}

We are back to the case of a single map contaminated by a characterised noise of spectrum:
\begin{equation}\label{eq:defNl}
  \mathcal{N}_\ell 
  = E {|\vec{W_\ell} \vec{a}_\lm^\mathrm{cont} |^2} 
  = 
  \left( \Acmb^\dag {\Rcont}^{-1}\Acmb \right)^{-1}
\end{equation}

If the residual $\vec{W_\ell} \vec{a}_\lm^\mathrm{cont}$ is modelled
as Gaussian, the single-map likelihood~(\ref{eq:loglikesingle}) can be
used.  

The same filter is used by \citet{2007PhRvD..75h3508A}. Assuming that the foreground contribution is perfectly known, the contaminant terms $\Rcont$ can be modelled as $\Rcont = \R[noise]_\ell + \R[fg]_\ell$. This approach thus permits to derive the actual level of contamination of the map in presence of known foregrounds, i.e. assuming that the covariance matrix of the foregrounds is known.

\subsection{Estimating $r$ in presence of unknown foregrounds with SMICA}
\label{sec:smica}

The two simplified approaches of sections \ref{sec:singlemap} and
\ref{sec:multimap} offer a way to estimate the impact of foregrounds
in a given mission, by comparing the sensitivity on $r$ obtained in
absence of foregrounds (from Eq.  \ref{eq:roughsigma} when $\modnl$
contains instrumental noise only), and the sensitivity achievable with
known foregrounds (when $\modnl$ contains the contribution of residual
contaminants as well, as obtained from Eq. \ref{eq:defNl} assuming
that the foreground correlation matrix is known).

A key issue, however, is that the solution {\emph {and the error bar}}
require the covariance matrix of foregrounds and noise to be
known.\footnote{The actual knowledge of the contaminant term is not strictly required to build the filter. It is required, however, to derive the contamination level of the filtered map.} Whereas the instrumental noise can be estimated accurately,
assuming prior knowledge of the covariance of the foregrounds to the
required precision is optimistic.

To deal with unknown foregrounds, we thus follow a different route
which considers a multi-map likelihood \citep{2003MNRAS.346.1089D}. If
all processes are modelled as Gaussian isotropic, then standard
computations yield:
\begin{equation}
  \label{eq:multi-like}
  -2 \loglik = \sum_\ell (2 \ell + 1) \fsky \KL{\R*_\ell}{\R_\ell} + \text{cst}
\end{equation}
where $\R*_\ell$ is the sample estimate of $\R_\ell$:
\begin{equation}
  \label{eq:rest}
  \R*_{\ell} =
  \frac{1}{2\ell+1}\frac1\fsky 
  \sum_{m=-\ell}^\ell \vec{a}_{l,m} \vec{a}_{l,m}^\dag 
\end{equation}
and where $\KL{\cdot}{\cdot}$ is a measure of mismatch between two
positive matrices given by:
\begin{equation}
  \label{eq:kullback}
  \KL{\R*}{\R} 
  =
  \frac{1}{2} \left[ \trace({\R}^{-1} \R*) - \logdet({\R}^{-1} \R*) - F \right]  
\end{equation}
Expression (\ref{eq:multi-like}) is nothing but the multi-map
extension of (\ref{eq:loglikesingle}).

If $\Rcont$ is known and fixed, then the
likelihood~(Eq. \ref{eq:multi-like}) depends only on the CMB angular
spectrum and can be shown to be equal (up to a constant) to
expression~\ref{eq:loglikesingle} with $C_\ell = r S_\ell$ and
$\modnl$ given by Eq.~\ref{eq:defNl}.
Thus this approach encompasses both the single map and filtered map approaches.


Unknown foreground contribution can be modelled as the mixed contribution of $D$ correlated sources:

\begin{equation}
  \label{eq:modelfg}
  \R[fg]_\ell = \tens{A} \tens{\Sigma}_\ell \tens{A}^\dag  
\end{equation}
 where $\tens{A}$ is a $F \times D$ mixing matrix and $\tens{\Sigma_\ell}$ is the $D \times D$ spectral covariance matrix of the sources. The model of the spectral covariance matrix of the observations is then:

\begin{displaymath}
  \R_\ell = r \shapecl \Acmb \Acmb^\dag + \tens{A} \tens{\Sigma}_\ell \tens{A}^\dag   + \R[noise]_\ell
\end{displaymath}
We then maximise the likelihood (\ref{eq:multi-like}) of the model with respect to $r$, $\tens{A}$ and $\tens{\Sigma}_\ell$.

We note that the foreground parameterisation in~Eq.\ref{eq:modelfg}
is redundant, as an invertible matrix can be exchanged between
\(\tens{A}\) and \(\tens{\Sigma}\), without modifying the actual value
of \R[fg]. The physical meaning of this is that the various
foregrounds are not identified and extracted individually, only their
mixed contribution is characterised. 

If we are interested in disentangling the foregrounds as well, e.g. to
separate synchrotron emission from dust emission, this degeneracy can
be lifted by making use of prior information to constrain, for
example, the mixing matrix. Our multi-dimensional model offers,
however, greater flexibility. Its main advantage is that no assumption
is made about the foreground physics. It is not specifically tailored
to perfectly match the model used in the simulation. Because of this,
it is generic enough to absorb variations in the properties of the
foregrounds, as will be seen later-on, but specific enough to preserve
identifiability in the separation of CMB from foreground emission. A
more complete discussion of the \smica\ method with flexible
components can be found in \citet{2008arXiv0803.1814C}.

A couple last details on \smica\ and its practical implementation are
of interest here. For numerical purposes, we actually divide the whole
$\ell$ range into $Q$ frequency bins $\domain=\lbrace
\ell^\text{min}_q, \cdots, \ell^\text{max}_q\rbrace$, and form the
binned versions of the empirical and true cross power-spectra:
\begin{equation}
\begin{split}
  \label{eq:rbin}
  \R*_{q} &= \frac{1}{w_q} \sum_{\ell \in \domain} \sum_{m=-\ell}^\ell \vec{a}_{l,m} \vec{a}_{l,m}^\dag\\
  \R_{q} &= \frac{1}{w_q}\sum_{\ell \in \domain} (2 \ell +1)\R_\ell
\end{split}
\end{equation}
where \(w_q\) is the number of modes in \domain. It is appropriate to
select the domains so that we can reasonably assume for each \(\ell
\in \domain , \R_\ell \approx \R_q\). This means that spectral bins
should be small enough to capture the variations of the power spectra.
In practice results are not too sensitive to the choice of the
spectral bin widths. Widths between 5 and 10 multipoles constitute a
good tradeoff.

Finally, we compute the Fisher information matrix $\tens I_{i,j}(\vec{\theta})$ deriving
from the maximised likelihood (\ref{eq:multi-like}) for the
parameter set $\vec{\theta} = \left(r, \tens{A}, \tens{\Sigma_1}, \cdots,
\tens{\Sigma_Q}\right)$:
\begin{equation}
  \label{eq:2}
  \tens I_{i,j}(\vec{\theta}) = \frac12 \sum_q w_q \trace \left(\frac{\partial \R_q(\vec{\theta})}{\partial \theta_i} {\R}_q^{-1} \frac{\partial \R_q(\vec\theta)}{\partial \theta_j} {\R}_q^{-1} \right)
\end{equation}
The lowest achievable variance of the $r$ estimate is obtained as the entry of the inverse of the FIM corresponding to the parameter $r$:

\begin{equation}
  \label{eq:errorsmica}
  \sigma_r^2 = \tens I^{-1}_{r,r}
\end{equation}

\section{Predicted results for various experimental designs}
\label{sec:results}

\begin{figure}
  \includegraphics{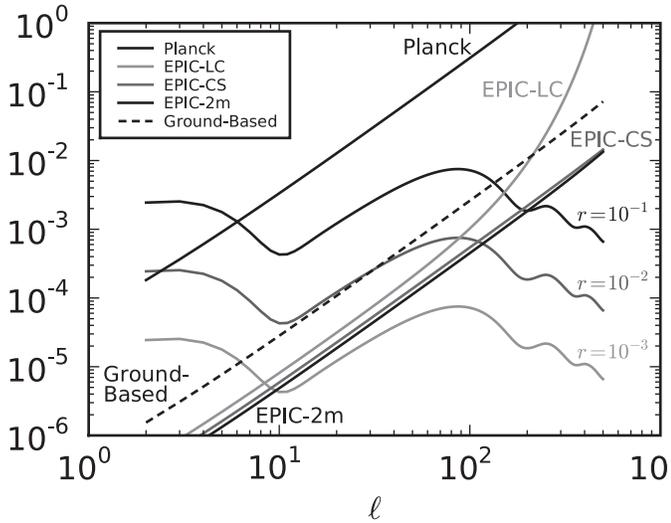}
  \caption{Noise spectra of various experimental designs compared to \Bmodes\ levels for $r = 0.1$, 0.01 and 0.001. When computing the equivalent multipole noise level for an experiment, we assume that only the central frequency channels contribute to the CMB measurement and that external channels are dedicated to foreground characterisation.}
  \label{fig:nl}
\end{figure}

We now turn to the numerical investigation of the impact of galactic foregrounds on the measurements of $r$ with the following experimental designs:
\begin{itemize}
\item The {\sc Planck} space mission, due for launch early 2009, which, although not originally planned for \Bmode\ physics, could provide a first detection if the tensor to scalar ratio $r$ is around $0.1$.
\item Various versions of the EPIC space mission, either low cost and low resolution (EPIC-LC), or more ambitious versions (EPIC-CS and EPIC-2m).
\item An ambitious (fictitious) ground-based experiment, based on the extrapolation of an existing design (the C\(\ell\)over experiment).
\item An alternative space mission, with sensitivity performances similar to the EPIC-CS space mission, but mapping only a small (and clean) patch of the sky, and referred as the `deep field mission'.
\end{itemize}
The characteristics of these instruments are summed-up in table \ref{tab:spec}, and Fig. \ref{fig:nl} illustrates their noise angular power spectra in polarisation.

\subsection{Pipeline}
\label{sec:pipeline}
\begin{figure}
  \includegraphics{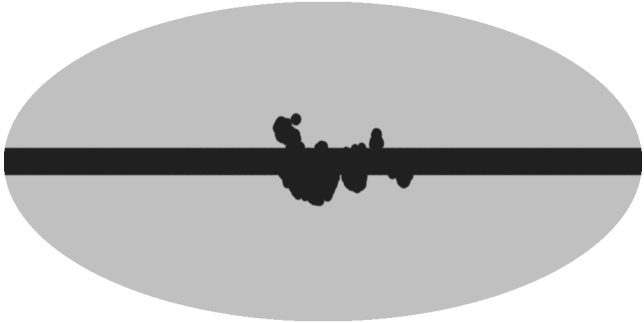}
  \caption{Analysis mask for EPIC B maps, smoothed with a $1^\circ$ apodisation window.}
  \label{fig:mask}
\end{figure}

For each of these experiments, we set up one or more simulation and analysis pipelines, which include, for each of them, the following main steps:
\begin{itemize}
\item Simulation of the sky emission for a given value of $r$ and a given foreground model, at the central frequencies and the resolution of the experiment.
\item Simulation of the experimental noise, assumed to be white, Gaussian and stationary.
\item Computation, for each of the resulting maps, of the coefficients of the spherical harmonic expansion of the \Bmodes\ $a^B_{\ell m}$
\item Synthesis from those coefficients of maps of B-type signal only.
\item For each experiment, a mask based on the B-modes level of the
  foregrounds is built to blank out the brightest features of the
  galactic emission (see Fig. \ref{fig:mask}). This mask is built with
  smooth edges to reduce mode-mixing in the pseudo-spectrum.
\item Statistics described in Equation \ref{eq:rbin} are built from the masked B maps.
\item The free parameters of the model described in Sect. \ref{sec:smica} are adjusted to fit these statistics. The shape of the CMB pseudo-spectrum that enters in the model, is computed using the mode-mixing matrix of the mask \citep{2002ApJ...567....2H}.
\item Error bars are derived from the Fisher information matrix of the model.
\end{itemize}

Some tuning of the pipeline is necessary for satisfactory foreground separation. The three main free parameters are the multipole range $[\ell_\text{min}, \ell_\text{max}]$, the dimension $D$ of the foreground component, and (for all-sky experiments) the size of the mask.

In practice we choose $\ell_\text{min}$ according to the sky coverage and $\ell_\text{max}$ according to the beam and the sensitivity.
The value of $D$ is selected by iterative increments until the goodness of fit (as measured from the \smica\ criterion on the data themselves, without knowledge of the input CMB and foregrounds) reaches its expectation. The mask is chosen in accordance to maximise the sky coverage for the picked value of $D$ (see appendix \ref{sec:d} for further discussion of the procedure).

For each experimental design and fiducial value of $r$ we compute
three kinds of error estimates which are recalled in Table
\ref{tab:results}.

Knowing the noise level and resolution of the instrument, we first
derive from Eq. \ref{eq:roughsigma} the error
$\sigma_r^\text{noise-only}$ set by the instrument sensitivity
assuming no foreground contamination in the covered part of the
sky. The global noise level of the instrument is given by $\modnl =
\left(\Acmb^\dag \Rcont^{-1} \Acmb \right)^{-1} $, where the only contribution to
\Rcont\ comes from the instrumental noise: $\Rcont =
\R[noise]_\ell = \diag\left(\frac{4\pi s_f^2}{B_\ell,f^2 t_s}\right)$.

In the same way, we also compute the error
$\sigma_r^\text{known-foreground}$ that would be obtained if
foreground contribution $\R[fg]$ to the covariance of the observations
was perfectly known, using $\Rcont = \R[noise]_\ell +
\R[fg]_\ell$. Here we assume that $\R[fg] = \R*[fg]$ where $\R*[fg]$
is the sample estimate of $\R[fg]$ computed from the simulated
foreground maps.
 
Finally, we compute the error $\sigma_r^\text{SMICA}$ given by the
Fisher information matrix of the model
(Eq. \ref{eq:errorsmica}).

In each case, we also decompose the FIM in the contribution from large
scale modes ($\ell \leq 20$) and the contribution from small scales
($\ell > 20$) to give indications of the relative importance of the
bump (due to reionisation) and the peak (at higher $\ell$) in the constraint of $r$.

We may notice that in some favourable cases (at low $\ell$, where the
foregrounds dominate), the error estimate given by \smica\ can be
slightly more optimistic than the estimate obtained using the actual empirical value of
the correlation matrix $\R*[fg]$. This reflects the fact that our
modelling hypothesis, which imposes to $\R[fg]$ to be of rank smaller
than $D$, is not perfectly verified in practice (see Appendix
\ref{sec:d} for further discussion of this hypothesis). The (small) difference (an error on the estimation of $\sigma_r$ when foregrounds are approximated by our model) has negligible impact on the conclusions of this work.

\begin{table*}
  \caption{Summary of experimental designs.}
  \label{tab:spec}
  \centering
  \begin{tabular}{lccccc}
    \hline
    \hline
    Experiment & frequency & beam FWHM & NET & $T_{obs}$ & sky coverage \\
    & (GHz) & (')& ($\mu K \sqrt{s} $) & (yr) &($f_{sky}$)\\
    \hline
    \multirow{2}{*}{PLANCK} & 30, 44, 70 & 33, 24, 14 &  96, 97, 97 & 1.2 & 1 \\
    &  100, 143, 217, 353 &  10, 7.1, 5, 5 &  41, 31, 51, 154 &  & \\
    \hline
    \multirow{2}{*}{EPIC-LC} & 30, 40, 60 & 155, 116, 77 & 28, 9.6, 5.3 & 2 & 1\\
    & 90, 135, 200, 300 & 52, 34, 23, 16 & 2.3, 2.2, 2.3, 3.8 &  & \\
    \hline
    \multirow{2}{*}{EPIC-CS} & 30, 45, 70, 100 & 15.5, 10.3, 6.6, 4.6 & 19, 8, 4.2, 3.2 & 4 & 1\\
    & 150, 220, 340, 500 & 3.1, 2.1, 1.4, 0.9 & 3.1, 5.2, 25, 210 &  & \\
    \hline
    \multirow{2}{*}{EPIC-2m} & 30, 45, 70, 100 & 26, 17, 11, 8 & 18, 7.6, 3.9, 3.0 & 4 & 1\\
    & 150, 220, 340, 500(,800) & 5, 3.5, 2.3, 1.5(, 0.9) & 2.8, 4.4, 20, 180(, 28k) & & \\
    \hline
    Ground-Based & 97, 150, 225 & 7.5, 5.5, 5.5 &  12, 18, 48 & 0.8 & 0.01 \\
    \hline
    \multirow{2}{*}{Deep field} & 30, 45, 70, 100 & 15.5, 10.3, 6.6, 4.6 & 19, 8, 4.2, 3.2 & 4 & 0.01\\
    & 150, 220, 340, 500 & 3.1, 2.1, 1.4, 0.9 & 3.1, 5.2, 25, 210 &  & \\
    \hline

  \end{tabular}
\end{table*}

\begin{table*}
  \caption{Error prediction for various experimental designs and fiducial $r$ values. Error bars from the columns noise-only and known foregrounds are derived from Eq.(\ref{eq:roughsigma}) assuming $\Rcont = \tens{R}^\text{noise}$ and $\Rcont =  \tens{R}^\text{noise} + \tens{R}^\text{fg}$ respectively. Error bars from the \smica\ column are obtained by the inversion of the FIM computed from the \smica\ model at the point of convergence of the algorithm as in Eq.(\ref{eq:errorsmica}). In all cases, large scale ($\ell \leq 20$) and small scale ($\ell > 20$) error bars are computed by decomposing the Fisher information between contribution from low and high multipoles. This allows for an estimation of respective contribution from the bump and the peak to the measurement. The $r^\text{est}$ column gives the estimated value at the convergence point in \smica. Detections at more than $4\sigma$ are bold-faced.}
  \label{tab:results}
  \centering
  \(
  \begin{array}{lc|ccc|ccc|ccc|c|cccc}
    \hline
    \hline
    & & \multicolumn{3}{|c|}{\text{noise-only}} & \multicolumn{3}{|c|}{\text{known foregrounds}} & \multicolumn{3}{|c|}{\text{\smica}} & \\
    \text{case} & r & \sigma_r/r & \sigma_r^{\ell \leq 20}/r & \sigma_r^{\ell > 20}/r & \sigma_r/r & \sigma_r^{\ell \leq 20}/r & \sigma_r^{\ell > 20}/r & \sigma_r/r & \sigma_r^{\ell \leq 20}/r & \sigma_r^{\ell > 20}/r & r^\text{est} & l_\text{min} - l_\text{max} & \fsky & D \footnote{Size of the galactic component used in the model of \smica.} \\
    \hline
    \multirow{2}{*}{\footnotesize PLANCK}  & 0.3 & 
0.075  & 0.17  & 0.084  &
0.1  & 0.2  & 0.12  &
0.15  & 0.22  & 0.2  & \mathbf{0.26} &   \multirow{2}{*}{2 - 130}  & \multirow{2}{*}{0.95} & \multirow{2}{*}{3} \\
     & 0.1 & 
0.17  & 0.25  & 0.22  &
0.23  & 0.34  & 0.32  &
0.29  & 0.34  & 0.55  & 0.086 &  \\

     \hline
    
     \multirow{2}{*}{\footnotesize EPIC-LC} 
      & 0.01 & 
0.019  & 0.084  & 0.019  &
0.05  & 0.18  & 0.053  &
0.079  & 0.18  & 0.1  & \mathbf{0.0098} &   \multirow{2}{*}{2 - 130}  & \multirow{2}{*}{0.86} & \multirow{2}{*}{4} \\
      & 0.001 & 
0.059  & 0.15  & 0.064  &
0.27  & 0.4  & 0.38  &
0.37  & 0.43  & 0.82  & 0.00088 &  \\
     \hline

     \multirow{2}{*}{\footnotesize EPIC-2m} 
      & 0.01 & 
0.016  & 0.083  & 0.016  &
0.027  & 0.12  & 0.027  &
0.032  & 0.11  & 0.036  & \mathbf{0.0096} &   \multirow{2}{*}{2 - 300}  & \multirow{2}{*}{0.87} & \multirow{2}{*}{4} \\
      & 0.001 & 
0.051  & 0.14  & 0.055  &
0.14  & 0.25  & 0.16  &
0.16  & 0.24  & 0.24  & \mathbf{0.001} &  \\
     \hline
     
     \multirow{2}{*}{\footnotesize EPIC-CS} 
      & 0.01 & 
0.017  & 0.084  & 0.017  &
0.029  & 0.12  & 0.03  &
0.036  & 0.11  & 0.041  & \mathbf{0.0096} &   \multirow{2}{*}{2 - 300}  & \multirow{2}{*}{0.87} & \multirow{2}{*}{4} \\
      & 0.001 & 
0.058  & 0.15  & 0.063  &
0.15  & 0.27  & 0.19  &
0.18  & 0.26  & 0.29  & \mathbf{0.00098} &  \\
     \hline
     \multirow{2}{*}{\footnotesize Ground-based}
      & 0.1 & 
0.083  & - & - &
0.15  & - & - &
0.24  & - & - & \mathbf{0.11} &   \multirow{2}{*}{50 - 300}  & \multirow{2}{*}{0.01} & \multirow{2}{*}{2} \\
      & 0.01 & 
0.18  & - & - &
0.8  & - & - &
1.6  & - & - & 0.018 &  \\
     \hline

     \text{\footnotesize Grnd-based+Planck} 
      & 0.01 & 
0.18  & - & - &
0.51  & - & - &
0.69  & - & - & 0.0065 &   \multirow{1}{*}{50 - 300}  & \multirow{1}{*}{0.01} & \multirow{1}{*}{2} \\
     \hline
     
     \text{\footnotesize Deep field mission} 
      & 0.001 & 
0.082  & - & - &
0.1  & - & - &
0.13  & - & - & \mathbf{0.00092} &   \multirow{1}{*}{50 - 300}  & \multirow{1}{*}{0.01} & \multirow{1}{*}{4} \\
     \hline

  \end{array}
  \)

\end{table*}

\subsection{Planck}
\label{sec:planck}
The Planck space mission will be the first all-sky experiment to give
sensitive measurements of the polarised sky in seven bands between 30
and 353~GHz. The noise level of this experiment being somewhat too
high for precise measurement of low values of $r$, we run our pipeline
for \(r = 0.1\) and \(0.3\). We predict a possible 3-sigma measurement
for \(r = 0.1\) using \smica\ (first lines in table
\ref{tab:results}). A comparison of the errors obtained from \smica,
with the prediction in absence of foreground contamination, and with
perfectly known foreground contribution, indicates that the error is
dominated by cosmic variance and noise, foregrounds contributing to a
degradation of the error of $\sim 30\%$ and uncertainties on
foregrounds for another increase around $30\%$ (for $r=0.1$).

Fig. \ref{fig:nl} hints that a good strategy to detect primordial
\Bmodes\ with Planck consists in detecting the reionisation bump below
$\ell = 10$, which requires the largest possible sky coverage. Even at
high latitude, a model using \(D=2\) fails to fit the galactic
emission, especially on large scales where the galactic signal is
above the noise. Setting \(D = 3\), however, gives a satisfactory fit
(as measured by the mismatch criterion) on 95 percent of the sky. It
is therefore our choice for Planck.
 
We also note that a significant part of the information is coming from
the reionisation bump ($\ell \leq 20$). The relative importance of the
bump increases for decreasing value of $r$, as a consequence of the
cosmic variance reduction. For a signal--to--noise ratio corresponding
roughly to the detection limit ($r = 0.1$), the stronger constraint is
given by the bump (Appendix \ref{sec:mm} gives further illustration of
the relative contribution of each multipole). This has two direct
consequences: the result is sensitive to the actual value of the
reionisation optical depth and to reionisation history (as
investigated by \citet{2008arXiv0804.0278C}), and the actual
capability of Planck to measure $r$ will depend on the level (and the
knowledge of) instrumental systematics on large scales.

Note that this numerical experiment estimates how well Planck can
measure $r$ in presence of foregrounds {\emph{from \Bmodes\ only}}.

\subsection{EPIC}
\label{sec:epic}

We perform a similar analysis for three possible designs of the EPIC
probe \citep{2008arXiv0805.4207B}. EPIC-LC and EPIC-CS correspond
respectively to the low cost and comprehensive solutions. EPIC-2m is
an alternate design which contains one extra high-frequency channel
(not considered in this study) dedicated to additional scientific
purposes besides CMB polarisation. We consider two values of $r$, 0.01
and 0.001. For all these three experiments, the analysis requires $D =
4$ for a reasonable fit, which is obtained using about 87\% of the
sky.

The two high resolution experiments provide measurements of \(r =
10^{-3}\) with a precision better than five sigma. For the lower
values of $r$, the error is dominated by foregrounds and their
presence degrades the sensitivity by a factor of 3, as witnessed by
the difference between $\sigma_r^\text{noise-only}$ and
$\sigma_r^\text{smica}$. However, while the difference between the
noise-only and the \smica\ result is a factor 4-6 for EPIC-LC, it is
only a factor about 2-3 for EPIC-CS and EPIC-2m. Increased
instrumental performance (in terms of frequency channels and
resolution) thus also allows for better subtraction of foreground
contamination.


For all experiments considered, the constraining power moves from
small scales to larger scale when $r$ decreases down to the detection
limit of the instrument.  In all cases, no information for the CMB is
coming from $\ell > 150$. Higher multipoles, however, are still giving
constraints on the foreground parameters, effectively improving the
component separation also on large scales.

\subsection{Small area experiments}
\label{sec:smallarea}
\subsubsection{Ground-based}
\label{sec:clover}

A different observation strategy for the measurement of \Bmodes\ is
adopted for ground-based experiments that cannot benefit from the
frequency and sky coverage of a space mission. Such experiments target
the detection of the first peak around \(\ell = 100\), by observing a
small but clean area (typically 1000 square-degrees) in few frequency
bands (2 or 3).

The test case we propose here is inspired from the announced
performances of C\(\ell\)over \citep{2008arXiv0805.3690N}. The
selected sky coverage is a 10 degree radius area centred on
$\text{lon} = 351^\circ$, $ \text{lat} = -56^\circ$ in galactic
coordinates. The region has been retained by the C\(\ell\)over team as
a tradeoff between several issues including, in particular, foreground and
atmospheric contamination.  According to our polarised galactic
foreground model, this also correspond to a reasonably clean part of
the sky (within 30\% of the cleanest).

The most interesting conclusion is that for $r =0.01$, although the
raw instrumental sensitivity (neglecting issues like E-B mixing due to
partial sky coverage) would allow a more than five sigma detection,
galactic foregrounds cannot be satisfactorily removed with the scheme
adopted here.

An interesting option would be to complement the measurement obtained
from the ground, with additional data as that of Planck, and extract
$r$ in a joint analysis of the two data sets.  To simply test this
possibility here, we complement the ground data set with a simulation
of the Planck measurements on the same area. This is equivalent to
extend the frequency range of the ground experiment with less
sensitive channels. We find a significant improvement of the error-bar
from $1.6 \cdot 10^{-2}$ to $0.69\cdot 10^{-2}$, showing that a joint
analysis can lead to improved component separation. The degradation of
sensitivity due to foreground remains however higher than for a fully
sensitive space mission (as witnessed by the following section).  This
last result is slightly pessimistic as we do not make use of the full
Planck data set but use it only to constrain foregrounds in the small
patch. However considering the ratio of sensitivity between the two
experiments, it is likely that there is little to gain by pushing the
joint analysis further.

\subsubsection{Deep field space mission}
\label{sec:hdf}
We may also question the usefulness of a full-sky observation strategy
for space-missions, and consider the possibility to spend the
whole observation time mapping deeper a small but clean region.

We investigate this alternative using an hypothetical experiment
sharing the sensitivity and frequency coverage of the EPIC-CS design,
and the sky coverage of the ground-based experiment.  Although the
absence of strong foreground emission may permit a design with a
reduced frequency coverage, we keep a design similar to EPIC-CS to
allow comparisons.  In addition, the relative failure of the
ground-based design to disentangle foregrounds indicates that the
frequency coverage cannot be freely cut even when looking in the
cleanest part of the sky.  In the same way, to allow straightforward
comparison with the ground-based case we stick to the same sky
coverage, although in principle, without atmospheric constraints,
slightly better sky areas could be selected.

In spite of the increased cosmic variance due to the small sky
coverage, the smaller foreground contribution allows our
harmonic-based foreground separation with \smica\ to achieve better
results with the `deep field' mission than with the full sky
experiment, when considering only diffuse galactic foreground.
However, this conclusion doesn't hold if lensing is considered as will
be seen in the following section.

We may also notice that, despite the lower level of foregrounds, the
higher precision of the measurement requires the same model complexity
($D = 4$) as for the full sky experiment to obtain a good fit.

We also recall that our processing pipeline does not exploit the
spatial variation of foreground intensity, and is, in this sense,
suboptimal, in particular for all-sky experiments. Thus, the results
presented for the full-sky experiment are bound to be slightly
pessimistic which tempers further the results of this comparison
between deep field and full sky mission. This is further discussed
below.
 
Finally, note that here we also neglect issues related to partial sky
coverage that would be unavoidable in this scheme.

\subsection{Comparisons}

\subsubsection{Impact of foregrounds: the ideal case}

As a first step, the impact of foregrounds on the capability to
measure $r$ with a given experiment, if foreground covariances are
known, is a measure of the adequacy of the experiment to deal with
foreground contamination. Figures for this comparison are computed
using equations \ref{eq:roughsigma} and \ref{eq:defNl}, and are given
in table \ref{tab:results} (first two sets of three columns).

The comparison shows that for some experiments, $\sigma_r/r$ in the
`noise-only' and the `known foregrounds' cases are very close. This is
the case for Planck and for the deep field mission. For these
experiments, if the second order statistics of the foregrounds are
known, galactic emission does not impact much the measurement. For
other experiments, the `known foregrounds' case is considerably worse
than the `noise-only' case. This happens, in particular, for a ground
based experiment when $r=0.01$, and for EPIC-LC.

If foreground contamination was Gaussian and stationary, and in
absence of priors on the CMB power spectrum, the linear filter of
equation \ref{eq:well-ideal} would be the optimal filter for CMB
reconstruction. The difference between $\sigma_r$ in the `noise-only'
and the `known foregrounds' cases would be a good measure of how much
the foregrounds hinder the measurement of $r$ with the experiment
considered. A large difference would indicate that the experimental
design (number of frequency channels and sensitivity in each of them)
is inadequate for `component separation'.

However, since foregrounds are neither Gaussian nor stationary, the
linear filter of equation \ref{eq:well-ideal} is not optimal. Even if
we restrict ourselves to linear solutions, the linear weights given to
the various channels should obviously depend on the local properties
of the foregrounds. Hence, nothing guarantees that we can not deal
better with the foregrounds than using a linear filter in harmonic
space. Assuming that the covariance matrix of the foregrounds is
known, the error in equation \ref{eq:roughsigma} with
$\mathcal{N}_\ell $ from equation \ref{eq:defNl} is a pessimistic
bound on the error on $r$. The only conclusion that can be drawn is
that the experiment does not allow effective component separation with
the implementation of a linear filter in harmonic space. There is,
however, no guarantee either that an other approach to component
separation would yield better results.

Hence, the comparison of the noise-only and known foregrounds cases
shown here gives an upper limit of the impact of foregrounds, if they
were known.

\subsubsection{Effectiveness of the blind approach}

Even if in some cases the linear filter of equation
\ref{eq:well-ideal} may not be fully optimal, it is for each mode
$\ell$ the best linear combination of observations in a set of
frequency channels, to reject maximally contamination from foregrounds
and noise, and minimise the error on $r$.  Other popular methods as
decorrelation in direct space, as the so-called `internal linear
combination' (ILC), and other linear combinations cannot do better,
unless they are implemented locally in both pixel and harmonic space
simultaneously, using for instance spherical needlets as in
\citet{2008arXiv0807.0773D}. Such localisation is not considered in
the present work.

Given this, the next question that arises is how well the spectral
covariance of the foreground contamination can be actually constrained
from the data, and how this uncertainty impact the measurement of
$r$. The answer to this question is obtained by comparing the second
and third sets of columns of table \ref{tab:results}.

In all cases, the difference between the results obtained assuming
perfect knowledge of the foreground residuals, and those obtained
after the blind estimation of the foreground covariances with \smica,
are within a factor of 2. For EPIC-2m and the deep field mission, the
difference between the two is small, which means that \smica\ allows
for component separation very effectively. For a ground based
experiment with three frequency channels, the difference is very
significant, which means that the data does not allow a good blind
component separation with \smica.

Comparing column set 1 (noise-only) and 3 (blind approach with \smica)
gives the overall impact of unknown galactic foregrounds on the
measurement of $r$ from \Bmodes\ with the various instruments
considered. For Planck, EPIC-2m, or a deep field mission with 8
frequency channels, the final error bar on $r$ is within a factor of 2
of what would be achievable without foregrounds. For EPIC-LC, or even
worse for a ground-based experiment, foregrounds are likely to impact
the outcome of the experiment quite significantly. For this reason,
EPIC-2m and the deep field mission seem to offer better perspectives
for measuring $r$ in presence of foregrounds.

\subsubsection{Full sky or deep field}

The numerical investigations performed here allow --to some extent--
to compare what can be achieved with our approach in two cases of sky
observation strategies with the same instrument. For EPIC-CS, it has
been assumed that the integration time is evenly spread on the entire
sky, and that 87\% of the sky is used to measure $r$. For the `deep
field' mission, 1\% of the sky only is observed with the same
instrument, with much better sensitivity per pixel (by a factor of
10).

Comparing $\sigma_r/r$ between the two in the noise-only case shows
that the full sky mission should perform better (by a factor 1.4) if
the impact of the foregrounds could be made to be negligible. This is
to be expected, as the cosmic or `sample' variance of the measurement
is smaller for larger sky coverage. After component separation
however, the comparison is in favour of the deep field mission, which
seems to perform better by a factor 1.4 also. The present work,
however, does not permit to conclude on what is the best strategy for
two reasons. First, this study concentrates on the impact of diffuse
galactic foregrounds which are not expected to be the limiting issue
of the deep field design.
And secondly, in the case of a deep field, the properties of the
(simulated) foreground emission are more homogeneous in the observed
area, and thus the harmonic filter of equation \ref{eq:well-ideal} is
close to optimal everywhere. For the full sky mission, however, the
filter is obtained as a compromise minimising the overall error $\ell$
by $\ell$, which is not likely to be the best everywhere on the
sky. Further work on component separation, making use of a localised
version of \smica, is needed to conclude on this issue. A preliminary
version of \smica\ in wavelet space is described in
\citet{2004astro.ph..7053M}, but applications to CMB polarisation and
full sky observations require specific developments.

\section{Discussion}
\label{sec:discussion}

The results presented in the previous section have been obtained using
a number of simplifying assumptions.

First of all, only galactic foregrounds (synchrotron and dust) are
considered. It has been assumed that other foregrounds (point sources,
lensing) can be dealt with independently, and thus will not impact
much the overall results.

Second, it is quite clear that the results may depend on details of
the galactic emission, which might be more complex
than what has been used in our simulations.

Third, most of our conclusions depend on the
accuracy of the determination of the error bars from the Fisher
information matrix. This method, however, only provides an
approximation, strictly valid only in the case of Gaussian processes
and noise.

Finally, the measurement of $r$ as performed here assumes a perfect
prediction (from other sources of information) of the shape of the BB
spectrum.

In this section, we discuss and quantify the impact of these
assumptions, in order to assess the robustness of our conclusions.

\subsection{Small scale contamination}

\subsubsection{Impact of lensing}
\label{sec:lensing}
Limitations on tensor mode detection due to lensing have been widely
investigated in the literature, and cleaning methods, based on the reconstruction of the lensed \Bmodes\ from
estimation of the lens potential and unlensed CMB E-modes,  have been
proposed
\citep{2002PhRvL..89a1303K,2003PhRvD..67d3001H,2003PhRvD..67l3507K,2006PhR...429....1L}.
However, limits on $r$ achievable after such `delensing' (if any) are
typically significantly lower than limits derived in
Sect. \ref{sec:results}, for which foregrounds and noise dominate the
error.
 
In order to check whether the presence of lensing can significantly
alter the detection limit, we proceed as follows: assuming no specific
reconstruction of the lens potential, we include lensing effects in
the simulation of the CMB (at the power spectrum level). The impact of
this on the second order statistics of the CMB is an additional
contribution to the CMB power spectrum. This extra term is taken into
account on the CMB model used in \smica. For this, we de--bias the CMB
\smica\ component from the (expectation value of) the lensing
contribution to the power-spectrum. The cosmic variance of the lensed
modes thus contributes as an extra `noise' which lowers the
sensitivity to the primordial signal, and reduces the range of
multipoles contributing significantly to the measurement.  We run this
lensing test case for the EPIC-CS and deep field mission.  Table
\ref{tab:lensing} shows a comparison of the constraints obtained with
and without lensing in the simulation for a fiducial value of $r =
0.001$. On large scales for EPIC-CS, lensing has negligible impact on
the measurement of $r$ (the difference between the two cases, actually
in favour of the case with lensing, is not significant on one single
run of the component separation). On small scales, the difference
becomes significant. Overall, $\sigma_r/r$ changes from 0.17 to 0.2,
not a very significant degradation of the measurement: lensing
produces a 15\% increase in the overall error estimate, the small
scale error (for $\ell > 20$) being most impacted. For the small
coverage mission, however, the large cosmic variance of the lensing
modes considerably hinder the detection.
 
\begin{table}[h]
  \centering
  \[
  \begin{array}{l|ccc|ccc}
    \hline
    \hline
    & \multicolumn{3}{c|}{\text{no lensing}} & \multicolumn{3}{c}{\text{lensing}} \\
    \text{Experiment} & \sigma_r/r & \sigma_r^{\ell \leq 20} /r & \sigma_r^{\ell > 20} /r & \sigma_r /r & \sigma_r^{\ell \leq 20} /r & \sigma_r^{\ell > 20} /r  \\
    \hline
    \text{EPIC-CS} &0.17 & 0.25 & 0.28  &0.2 & 0.24 & 0.36  \\
    \text{Deep field} &0.13 & -& - &1.1 & -& - \\
    \hline
  \end{array}
  \]
  \caption{Comparison of the constraints on $r$ with and without lensing (here $r=0.001$).}
  \label{tab:lensing}
\end{table}

Thus, at this level of $r$, if the reionisation bump is satisfactorily
measured, the difference is perceptible but not very
significant. Hence, lensing is not the major source of error for a
full-sky experiment measuring $r$. It becomes however a potential
problem for a small coverage experiment targeting the measurement of
the recombination bump. Such a strategy would thus require efficient
`delensing'. Indications that `delensing' can be performed even in
presence of foregrounds in the case of a low noise and high resolution
experiment can be found in \citet{2008arXiv0811.3916S}. However, a
complete investigation of this case, accounting for all the complexity
(diffuse foregrounds, point sources, lensing, modes-mixing effects),
would be needed to conclude on the validity of a deep-field strategy.

\subsubsection{Impact of extra-galactic sources}
\label{sec:radiosource}
Although largely sub-dominant on scales larger than 1 degree,
extra-galactic sources, in particular radio-sources, are expected to
be the worst contaminant on small scales (see
e.g. \citet{tucci/etal:2004,Pierpaoli04}).

Obviously, the strongest point sources are known, or (for most of
them) will be detected by Planck. Their polarisation can be measured
either by the \Bmode\ experiment itself, or by dedicated follow-up.
We make the assumption that point sources brighter than 500~mJy in
temperature (around 6000 sources) are detected, and that their
polarised emission is subtracted from the polarisation observations.
We stress that 500~mJy is a conservative assumption as Planck is
expected to have better detection thresholds.

The present level of knowledge about point sources does not allow a
very accurate modelling of the contribution to the power spectra of
the remaining point sources (those not subtracted by the 500~mJy
cut). For this reason we investigate their impact in two extreme
cases: perfect modelling of their contribution to the power-spectra
(`ideal' case), and no specific modelling at all (`no-model' case).
Results of a \smica\ run for both assumptions are compared to what is
obtained in total absence of point sources (`no-ps' case), and are
summarised in table \ref{tab:ps}.

\begin{table*}
  \caption{EPIC-CS measurement for three point sources cases. For the reference case `no-ps', point sources have neither been including in the simulation, nor taken into account in the modelling. The `ideal' case presents an optimistic situation where the exact contribution of the point sources put in the simulation has been used to build the model. The `no-model' case is a pessimistic situation where no effort has been made to model the point sources contribution.}
  \label{tab:ps}
  \centering
  \[
  \begin{array}{c|ccc|ccc}
    \hline
    \hline
      r & r^\text{no-ps} & r^\text{ideal} & r^\text{no-model} & \sigma_r^\text{no-ps} & \sigma_r^\text{ideal} & \sigma_r^\text{no-model} \\
      
       0.001 & 1.07 \cdot 10^{-3} & 1.04 \cdot 10^{-3} & 2.00 \cdot 10^{-3} & 1.84 \cdot 10^{-4} & 1.91 \cdot 10^{-4} & 2.49 \cdot 10^{-4} \\
       \hline
  \end{array}
  \]
\end{table*}

\begin{figure}
  \includegraphics{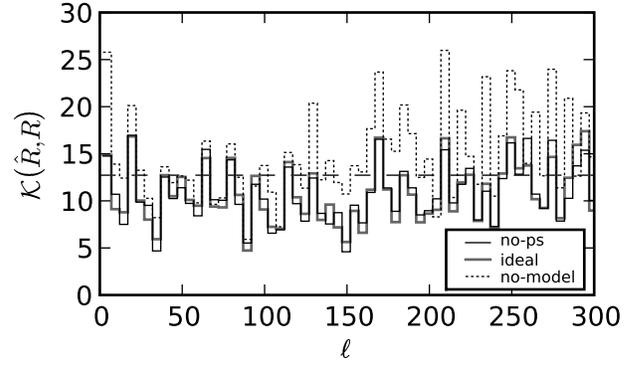}
  \caption{Goodness--of--fit for the three point sources cases. For the reference case `no-ps', point sources have neither been including in the simulation, nor taken into account in the modelling. The mismatch criterion wander around its expectation value (horizontal dashed line). The `no-model' case is a pessimistic situation where no effort has been made to model the point sources contribution, yielding a net increase of the mismatch criterion. The `ideal' case presents an optimistic situation where the exact contribution of the simulated point sources has been used to build the model. This perfect modelling restore the goodness--of--fit of the no-ps case.}
  \label{fig:mmps}

\end{figure}

The bottom line of this investigation is that modelling properly the
point sources statistical contribution is necessary to measure
$r=0.001$.  An insufficient model results in a biased estimator: for
EPIC-CS the estimated $r$ is two times larger than expected, with a
difference incompatible with the error bar, in spite of an increased
standard deviation ($\sigma_r$ increased by +30\% for $r=0.001$).

An ideal model restores the goodness of fit of the no-ps case and
suppresses the bias of the estimator. Still, the presence of point
sources increases the variance of the measurement of $r$. In our
experiment, the effect is not truly significant ($\sigma_r$ shifting
from 1.84 to $1.91 \cdot 10^{-4}$).

Figure \ref{fig:mmps} shows the mismatch criterion (from
Eq. \ref{eq:kullback}, using covariance matrixes binned in $\ell$) in
the three cases. When no specific model of the point source
contribution is used, some of their emission is nonetheless absorbed
by the \smica\ `galactic' component, which adjusts itself (via the
values of its maximum likelihood parameters) to represent best the
total foreground emission. The remaining part is responsible for the
increase of the mismatch at high $\ell$. At the same time, the
galactic estimation is twisted by the presence of point sources. This
slightly increases the mismatch on large scales.

\subsection{Galactic foregrounds uncertainties}

We now investigate the impact on the above results of modifying
somewhat the galactic emission. In particular, we check whether a
space dependant curvature of the synchrotron spectral index, and
modifications of the dust angular power spectrum, significantly change
the error bars on $r$ obtained in the previous section.

\subsubsection{Impact of synchrotron curvature}
\label{sec:synccurv}
\begin{figure*}
  \includegraphics{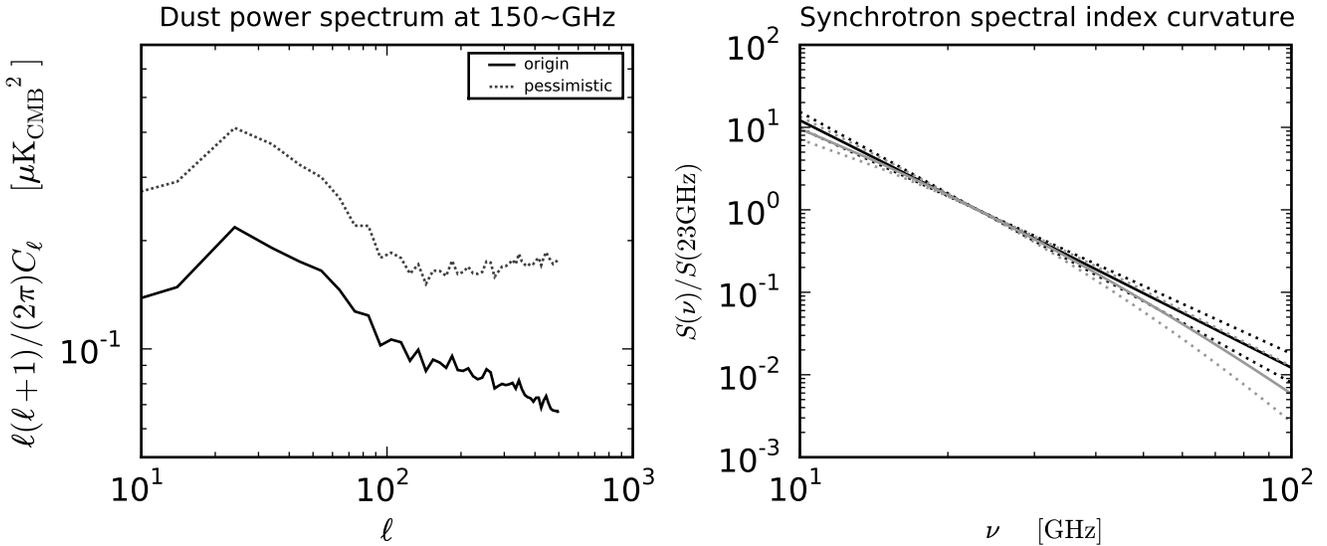}
  \caption{Variations of the galactic foregrounds model. The left panel shows the difference between the default power spectrum of dust polarisation \Bmodes\ at 150~GHz as modelled by the PSM (solid curve) and a model assuming pessimistic values for the overall level and power spectrum index (dotted curve). The right panel shows the dispersion of the synchrotron spectral index for the PSM model (in black) and the curved model (in gray). Solid lines present the frequency scaling for the mean values of the spectral index and dotted lines for its extremal values.}
  \label{fig:fgvar}
\end{figure*}

As mentioned earlier on, the synchrotron emission law may not be
perfectly described as a single power law per pixel, with a constant
spectral index across frequencies. Steepening of the spectral index is
expected in the frequency range of interest. As this variation is
related to the aging of cosmic rays, it should vary on the sky.
Hence, the next level of sophistication in modelling synchrotron
emission makes use of a (random) template map \(C(\spos)\) to model
the curvature of the synchrotron spectral index.  We then produce
simulated synchrotron maps as:
\begin{equation}
S_{\nu}^X(\spos) = S^X_{\nu_0}(\spos) \left(\cfrac{\nu}{\nu_0}\right)^{\beta_s(\spos)+ \alpha C(\spos) \log(\nu / \nu_1)}
\end{equation}
where \(\alpha\) is a free parameter which allows to modulate the
amplitude of the effect (as compared to equation
\ref{eq:syncelaw}). The right panel of figure \ref{fig:fgvar}
illustrates the impact of the steepening on the synchrotron frequency
scaling.

We now investigate whether such a modified synchrotron changes the
accuracy with which $r$ can be measured. We decide, for illustrative
purposes, to perform the comparison for EPIC-2m, and for $r=0.001$.
Everything else, regarding the other emissions and the foreground
model in \smica, remains unchanged. Table \ref{tab:alpha} shows the
results of this study in terms of goodness of fit and influence on the
\(r\) estimate. We observe no significant effect, which indicates that
the foreground emission model of Eq.(\ref{eq:modelfg}) is flexible
enough to accommodate the variation of the synchrotron modelling. Even
if we cannot test all possible deviation from the baseline PSM model,
robustness against running of the spectral index remains a good
indication that results are not overly model dependent.

\begin{table}[h]
  \caption{Influence of the running of the synchrotron spectral index on component separation in term of goodness of fit and \(r\) estimates. The study is conducted for the EPIC-2m design, for varying amplitude \(\alpha\) of the running of the spectral index. No significant variation of the $r$ estimate nor of the likelihood of the model is noticed for $\alpha$ remaining in the range allowed by observations.}
  \label{tab:alpha}
  \centering
  \[
  \begin{array}{cc|c|c|c}
    \hline
    \hline
    r & \sigma_r & \alpha & r & -\loglik \\
    \hline
    \multirow{3}{*}{\( 0.001 \)} & \multirow{3}{*}{\( 1.8 \cdot 10^{-4} \)}
    &0 & 9.78 \cdot 10^{-4} & 11.6 \\
    & &1 & 9.62 \cdot 10^{-4} & 11.5 \\
    & &3 & 1.06 \cdot 10^{-3} & 11.7 \\
    \hline
  \end{array}
  \]
\end{table}

\subsubsection{Level and power spectrum of dust emission}

Similarly, we now vary the model of dust emission and check how the
main results of section \ref{sec:results} are modified.  Measurements
give some constraints on dust emission on large scales, but smaller
scales remain mostly unconstrained. Hence, we consider here a
pessimistic extreme in which we multiply the large scale level of the
dust by a factor of two, and flatten the power spectrum from a nominal
index of -2.5 to -1.9.  The power spectra corresponding to these two
cases are shown in Fig. \ref{fig:fgvar} (left panel).

\begin{table}[h]
  \caption{Influence of dust polarisation level on component separation. We comparison results for a pessimistic (17\% intrinsic polarisation fraction, flat spectrum) and standard (12\% polarisation fraction) model of the dust emission.}
  \label{tab:dustflat}
  \centering
  \(
  \begin{array}{cc|cccc}
    \hline
    \hline
  \text{Experiments} & r & r^\text{origin} & r^\text{pessim} & \sigma_r^\text{origin} & \sigma_r^\text{pessim} \\
  \hline
  \text{Ground-based} & 0.01 & 1.84 \cdot 10^{-2} & 1.69 \cdot 10^{-2} & 1.62 \cdot 10^{-2} & 1.62 \cdot 10^{-2}\\
  \text{EPIC-2m} & 0.001 & 8.77 \cdot 10^{-4} & 8.77 \cdot 10^{-4} & 3.68 \cdot 10^{-4} & 3.61 \cdot 10^{-4}\\
  \hline
  \end{array}
  \)
  
\end{table}

Running the same component separation pipeline for the ground based
and the EPIC-2m experiments at their detection limit, we find only
marginal changes in the measured values of $r$ (see table
\ref{tab:dustflat}). This result can be interpreted in the following
way: as the noise of the experiment remains unchanged, the increased
signal-to-noise ratio allows for a better constraint of the dust
parameters. Component separation effectiveness depends mainly on the
coherence of the component, rather than on its overall level.

\subsection{Error bar accuracy}
\label{sec:accuracy}

Estimates of the error derived from the FIM (Eq. \ref{eq:errorsmica})
are expected to be meaningful only if the model leading to the
likelihood (Eq. \ref{eq:multi-like}) holds. In particular we assume
that processes can be modelled as Gaussian.

We first note that the FIM errors are reasonably compatible with the
difference between input and measured $r$ values, which gives
confidence that these error estimates are not obviously wrong.
Nonetheless, we investigate this issue further, using Monte-Carlo
studies to obtain comparative estimates of errors, with the EPIC-CS
design.  Table \ref{tab:mc} gives, for two values of $r$ and for 100
runs of the \smica\ pipeline in each case, the average recovered value
of $r$, the average error as estimated from the Fisher matrix $\langle
\sigma_r^\text{FISHER} \rangle$, and the standard deviation
$\sigma_r^\text{MC}$ of the measured values of $r$.

For each of the Monte-Carlo runs, a new realization of CMB and noise
is generated. Simulated galactic foregrounds, however, remain
unchanged.

Results show that the FIM approximation give estimates of the error in
very good agreement with the MC result.
Hence, the FIM estimate looks good enough for the purpose of the present
paper, given the number of other potential sources of error and the
computational cost of Monte-Carlo studies.

The Monte-Carlo study also allows to investigate the existence of a
bias. For an input tensor to scalar ratio of $0.01$, we observe that
the measured value of $r$ seems to be systematically low, with an
average of $9.91 \cdot 10^{-3}$.  This we interpret as resulting from
a slight over-fitting of the data.  Still this small bias
doesn't dominate the error and we are more interested in noise
dominated regime. The overall conclusion of this investigation of
error bars is that the errors estimated by the FIM are reasonably
representative of the measurement error.

\begin{table}
  \caption{Monte-Carlo analysis for error bars of the EPIC-CS experiment for 2 representative values of $r$. Sample mean and variance are obtained on 100 realizations of noise and CMB. $\langle r \rangle$ denotes the average recovered value of
$r$, $\langle \sigma_r^\text{FISHER} \rangle$ the average error as estimated from the Fisher matrix, and $\sigma_r^\text{MC}$ the standard deviation of the measured values of $r$}
  \label{tab:mc}
  \centering
  \[
  \begin{array}{c|ccc}
    \hline
    \hline
    r & \langle r \rangle & \langle \sigma_r^\text{FISHER} \rangle & \sigma_r^\text{MC} \\
    \hline
    0.01 & 9.91 \cdot 10^{-3} & 3.59 \cdot 10^{-4} & 3.49 \cdot 10^{-4} \\
    0.001 & 1.05 \cdot 10^{-3} & 1.84 \cdot 10^{-4} & 1.84 \cdot 10^{-4} \\
    \hline
  \end{array}
  \]
\end{table}

\subsection{Other cosmological parameters}
\label{sec:tau}

The main conclusions of this study are mostly independent of the value
of all cosmological parameters except $\tau$. Within present
uncertainties indeed, only the value of the reionisation optical depth
\(\tau\), which drives the amplitude and position of the reionisation
bump, is critical for our estimations \citep{colombo2008}. Lower
$\tau$ means less accurate measurement of $r$, and higher $\tau$
better measurement of $r$.  Here we choose a rather conservative value
of \(\tau = 0.07\) in agreement with the last measurements from WMAP
\citep{2008arXiv0803.0586D,2008arXiv0811.4280D}. The value of $\tau$,
however, should affect mainly low resolution and noisy experiments,
for which most of the information comes from the lowest frequency
`reionisation' bump in the \Bmode\ spectrum.

Another issue is that we assume the value of $\tau$ and $n_t$ (and, to
a less extent, the value of all other cosmological parameters) to be
perfectly known (setting the shape of the \Bmode\ power spectrum). In
fact, uncertainties on all cosmological parameters imply that the
shape will be known only approximately, and within a certain
framework. Such uncertainties will have to be taken into account in
the analysis of a real-life data set. Our \smica\ pipeline can be
adapted to do this, provided we know the uncertainties on the
cosmological parameter set. A Monte-Carlo approach, in which we
assume, for each \smica\ run, a \Bmode\ power spectrum from one of the
possible cosmological parameter sets, will permit to propagate the
uncertainties onto the measurement of $r$. We expect, however, that
this additional error will be significantly smaller than that due to
the experimental noise.

\section{Conclusion}
\label{sec:conclusion}

In this paper, we presented an investigation of the impact of
foregrounds on the measurement of the tensor to scalar ratio of
primordial perturbations. The measurement of $r$ is based on the
(simulated) observation of the \Bmode\ polarisation of the Cosmic
Microwave Background by various instruments, either in preparation or
planned for the future: the Planck space mission, a ground-based
experiment of the type of C$\ell$over, and several versions of a
possible dedicated space mission.

Foreground contamination is modelled and simulated using the present
development version (v1.6.4) of the Planck Sky Model (PSM).  Our main
analysis considers the contribution from diffuse polarised emission
(from the galactic interstellar medium modelled as a mixture of
synchrotron emission and thermal emission from dust) and from
instrumental noise. The impact of more complicated galactic foreground
emission, and of point sources and lensing, is investigated in a
second step.

Our approach uses the \smica\ component separation method on maps of
\Bmodes\ alone.
The method is robust with respect to specifics of foreground emission,
because it does not rely on an accurate representation of foreground
properties.  That last point is demonstrated by varying the input
foreground sky, and comparing results obtained with different inputs,
without changing the analysis pipeline.

It is shown that for $r$ at the level of $r \simeq 0.1$, Planck could
make a meaningful ($3\sigma$) detection from \Bmodes\ alone.  The
final sensitivity of Planck for measuring $r$ may be better than what
is achieved here, as a significant part of the constraining power on
$r$ should also come from EE/TE for high $r$. This has not been
investigated in the present paper, which is more focussed on the
measurement of low values of $r$ (not achievable with Planck).  With
the various EPIC mission designs, one could achieve detections at
levels of 4-8$\sigma$ for $r=10^{-3}$.

For full--sky, multi-frequency space missions, dealing with foregrounds
in harmonic space results in a loss of sensitivity by a factor 3 to
4, as compared to what would be achievable without foregrounds,
even if the covariance of foreground contaminants is known.  The
\smica\ pipeline allows to achieve performances almost as good (within
a factor 1.5), which demonstrates the effectiveness of the blind
approach, but is still significantly worse (factor 3-5) than if there were no
foregrounds at all.  The loss of sensitivity is probably due in part
to insufficient localisation in pixel space, which results in
sub--optimality of the estimator.  This could (at least in principle)
be improved with a localised processing.

For the most ambitious EPIC space mission, we find that our main
conclusions are not modified significantly when taking into account
the contamination of primordial \Bmodes\ by extra-galactic point
sources, by gravitational lensing, or when simulating a more
complicated galactic emission.
In contrast, we find that the measurement of $r$ from the ground with
few frequency channels can be severely compromised by foregrounds,
even in clean sky regions.

The joint analysis of such ground-based data together with those from
less sensitive experiments covering a wider frequency range, as the
Planck data, permits to improve the constraints on $r$. Still, the
result from a combined analysis of Planck and of a small patch
observed from the ground at few frequencies cannot match what is
obtained using sensitive measurements on the whole frequency range.

This makes a strong case for sensitive multi-frequency observations,
and thus probably also for a space mission, as observations from the
ground are severely limited (in frequency coverage) by atmospheric
absorption and emission.  This conclusion is further supported by the
fact that a space mission mapping the same clean region (about 1\% of
the sky), but with the full frequency range allowed by the absence of
atmosphere, makes it possible to deal with diffuse foregrounds very
efficiently.

Such a deep field mission would, in that respect, outperform a
comparable full-sky experiment. The results obtained in the present
study, however, do not permit to conclude whether a full sky or a deep
field mission would ultimately perform better. A strategy based on the
observation of a small patch seems to offer better prospects for
measuring $r$ with an harmonic--space based version of \smica, but
also seems to be more impacted by small scale contamination than
all-sky experiments, and is in particular quite sensitive to the
lensing effect. Further developments of the component separation
pipeline could improve the processing of both types of datasets.

As a final comment, we would like to emphasise that the present study
is, to our knowledge, the first one which designs, implements
effectively, and tests thoroughly on numerous simulations a component
separation method for measuring $r$ with CMB \Bmodes\ without relying
much on a physical model of foreground emission. The method is shown
to be robust against complicated foregrounds (pixel-dependent and
running synchrotron spectral index, multi-template dust emission,
polarised point sources and lensing). It is also shown to provide
reliable errors bars on $r$ by comparing analytical error bars (from
the FIM) to estimates obtained from Monte-Carlo simulations. Although
more work is needed for the optimal design of the next \Bmode\
experiment, our results demonstrate that foregrounds can be handled
quite effectively, making possible the measurement of $r$ down to
values of 0.001 or better, at the 5-6$\sigma$ level.

Certainly, next steps will require fully taking into account small
scale contaminants, partial sky coverage effects, and probably some
instrumental effects in addition to diffuse foregrounds. For this
level of detail, however, it would be mandatory to refine as well the
diffuse foreground model, using upcoming sensitive observations of the
sky in the frequency range of interest and on both large and small
scales. Such data will become available soon with the forthcoming
Planck mission.

\begin{acknowledgements} 
  The authors acknowledge the use of the Planck Sky Model, developed
  by the Component Separation Working Group (WG2) of the Planck
  Collaboration.  The HEALpix package was used for the derivation of
  some of the results presented in this paper.  MB and EP were
  partially supported by JPL SURP award 1314616 for this work.  MB
  would like to thank USC for hospitality during the Spring 2008 and
  Marc-Antoine Miville-Desch\^ enes for sharing his expertise on
  galactic foreground modelling.  EP is an NSF--ADVANCE fellow
  (AST--0649899) also supported by NASA grant NNX07AH59G and Planck
  subcontract 1290790.  JD, JFC and MLJ were partially supported by
  the ACI `Astro-Map' grant of the French ministry of research to
  develop the \smica\ component separation package used here.

\end{acknowledgements}
\bibliographystyle{aa}
\bibliography{bmodes} 

\begin{thebibliography}{67}
\expandafter\ifx\csname natexlab\endcsname\relax\def\natexlab#1{#1}\fi

\bibitem[{{Amblard} {et~al.}(2007){Amblard}, {Cooray}, \&
  {Kaplinghat}}]{2007PhRvD..75h3508A}
{Amblard}, A., {Cooray}, A., \& {Kaplinghat}, M. 2007, \prd, 75, 083508

\bibitem[{{Audit} \& {Simmons}(1999)}]{1999MNRAS.305L..27A}
{Audit}, E. \& {Simmons}, J.~F.~L. 1999, \mnras, 305, L27

\bibitem[{{Aumont} \& {Mac{\'{\i}}as-P{\'e}rez}(2007)}]{2007MNRAS.376..739A}
{Aumont}, J. \& {Mac{\'{\i}}as-P{\'e}rez}, J.~F. 2007, \mnras, 376, 739

\bibitem[{{Battistelli} {et~al.}(2006){Battistelli}, {Rebolo},
  {Rubi{\~n}o-Mart{\'{\i}}n}, {Hildebrandt}, {Watson}, {Guti{\'e}rrez}, \&
  {Hoyland}}]{2006ApJ...645L.141B}
{Battistelli}, E.~S., {Rebolo}, R., {Rubi{\~n}o-Mart{\'{\i}}n}, J.~A., {et~al.}
  2006, \apjl, 645, L141

\bibitem[{{Baumann} \& {Peiris}(2008)}]{2008arXiv0810.3022B}
{Baumann}, D. \& {Peiris}, H.~V. 2008, ArXiv e-prints

\bibitem[{{Bennett} {et~al.}(2003){Bennett}, {Hill}, {Hinshaw}, {Nolta},
  {Odegard}, {Page}, {Spergel}, {Weiland}, {Wright}, {Halpern}, {Jarosik},
  {Kogut}, {Limon}, {Meyer}, {Tucker}, \& {Wollack}}]{2003ApJS..148...97B}
{Bennett}, C.~L., {Hill}, R.~S., {Hinshaw}, G., {et~al.} 2003, \apjs, 148, 97

\bibitem[{{Beno{\^i}t} {et~al.}(2004){Beno{\^i}t}, {Ade}, {Amblard}, {Ansari},
  {Aubourg}, {Bargot}, {Bartlett}, {Bernard}, {Bhatia}, {Blanchard}, {Bock},
  {Boscaleri}, {Bouchet}, {Bourrachot}, {Camus}, {Couchot}, {de Bernardis},
  {Delabrouille}, {D{\'e}sert}, {Dor{\'e}}, {Douspis}, {Dumoulin}, {Dupac},
  {Filliatre}, {Fosalba}, {Ganga}, {Gannaway}, {Gautier}, {Giard},
  {Giraud-H{\'e}raud}, {Gispert}, {Guglielmi}, {Hamilton}, {Hanany},
  {Henrot-Versill{\'e}}, {Kaplan}, {Lagache}, {Lamarre}, {Lange},
  {Mac{\'{\i}}as-P{\'e}rez}, {Madet}, {Maffei}, {Magneville}, {Marrone},
  {Masi}, {Mayet}, {Murphy}, {Naraghi}, {Nati}, {Patanchon}, {Perrin}, {Piat},
  {Ponthieu}, {Prunet}, {Puget}, {Renault}, {Rosset}, {Santos}, {Starobinsky},
  {Strukov}, {Sudiwala}, {Teyssier}, {Tristram}, {Tucker}, {Vanel}, {Vibert},
  {Wakui}, \& {Yvon}}]{2004A&A...424..571B}
{Beno{\^i}t}, A., {Ade}, P., {Amblard}, A., {et~al.} 2004, \aap, 424, 571

\bibitem[{{Bock} {et~al.}(2008){Bock}, {Cooray}, {Hanany}, {Keating}, {Lee},
  {Matsumura}, {Milligan}, {Ponthieu}, {Renbarger}, \&
  {Tran}}]{2008arXiv0805.4207B}
{Bock}, J., {Cooray}, A., {Hanany}, S., {et~al.} 2008, ArXiv e-prints

\bibitem[{{Cardoso} {et~al.}(2008){Cardoso}, {Le Jeune}, {Delabrouille},
  {Betoule}, \& {Patanchon}}]{2008arXiv0803.1814C}
{Cardoso}, J.-F., {Le Jeune}, M., {Delabrouille}, J., {Betoule}, M., \&
  {Patanchon}, G. 2008, ArXiv e-prints, 803

\bibitem[{{Challinor} {et~al.}(2003){Challinor}, {Chon}, {Hivon}, {Prunet}, \&
  {Szapudi}}]{2003NewAR..47..995C}
{Challinor}, A., {Chon}, G., {Hivon}, E., {Prunet}, S., \& {Szapudi}, I. 2003,
  New Astronomy Review, 47, 995

\bibitem[{{Colombo} \& {Pierpaoli}(2008)}]{2008arXiv0804.0278C}
{Colombo}, L.~P.~L. \& {Pierpaoli}, E. 2008, ArXiv e-prints

\bibitem[{{Colombo} {et~al.}(2008){Colombo}, {Pierpaoli}, \&
  {Pritchard}}]{colombo2008}
{Colombo}, L.~P.~L., {Pierpaoli}, E., \& {Pritchard}, J.~R. 2008, ArXiv
  e-prints

\bibitem[{{de Oliveira-Costa} {et~al.}(2004){de Oliveira-Costa}, {Tegmark},
  {Davies}, {Guti{\'e}rrez}, {Lasenby}, {Rebolo}, \&
  {Watson}}]{2004ApJ...606L..89D}
{de Oliveira-Costa}, A., {Tegmark}, M., {Davies}, R.~D., {et~al.} 2004, \apjl,
  606, L89

\bibitem[{{de Zotti} {et~al.}(2005){de Zotti}, {Ricci}, {Mesa}, {Silva},
  {Mazzotta}, {Toffolatti}, \& {Gonz{\'a}lez-Nuevo}}]{2005A&A...431..893D}
{de Zotti}, G., {Ricci}, R., {Mesa}, D., {et~al.} 2005, \aap, 431, 893

\bibitem[{{Delabrouille} {et~al.}(2008){Delabrouille}, {Cardoso}, {Le Jeune},
  {Betoule}, {Fay}, \& {Guilloux}}]{2008arXiv0807.0773D}
{Delabrouille}, J., {Cardoso}, J.~., {Le Jeune}, M., {et~al.} 2008, ArXiv
  e-prints

\bibitem[{{Delabrouille} {et~al.}(2003){Delabrouille}, {Cardoso}, \&
  {Patanchon}}]{2003MNRAS.346.1089D}
{Delabrouille}, J., {Cardoso}, J.-F., \& {Patanchon}, G. 2003, \mnras, 346,
  1089

\bibitem[{{Delabrouille et al.}(2009)}]{Delabrouille09}
{Delabrouille et al.} 2009, in preparation

\bibitem[{{D\'esert} {et~al.}(1990){D\'esert}, {Boulanger}, \&
  {Puget}}]{1990A&A...237..215D}
{D\'esert}, F.-X., {Boulanger}, F., \& {Puget}, J.~L. 1990, \aap, 237, 215

\bibitem[{{Draine} \& {Fraisse}(2008)}]{2008arXiv0809.2094D}
{Draine}, B.~T. \& {Fraisse}, A.~A. 2008, ArXiv e-prints

\bibitem[{{Draine} \& {Lazarian}(1998)}]{1998ApJ...508..157D}
{Draine}, B.~T. \& {Lazarian}, A. 1998, \apj, 508, 157

\bibitem[{{Dunkley} {et~al.}(2008{\natexlab{a}}){Dunkley}, {Komatsu}, {Nolta},
  {Spergel}, {Larson}, {Hinshaw}, {Page}, {Bennett}, {Gold}, {Jarosik},
  {Weiland}, {Halpern}, {Hill}, {Kogut}, {Limon}, {Meyer}, {Tucker}, {Wollack},
  \& {Wright}}]{2008arXiv0803.0586D}
{Dunkley}, J., {Komatsu}, E., {Nolta}, M.~R., {et~al.} 2008{\natexlab{a}},
  ArXiv e-prints, 803

\bibitem[{{Dunkley} {et~al.}(2008{\natexlab{b}}){Dunkley}, {Spergel},
  {Komatsu}, {Hinshaw}, {Larson}, {Nolta}, {Odegard}, {Page}, {Bennett},
  {Gold}, {Hill}, {Jarosik}, {Weiland}, {Halpern}, {Kogut}, {Limon}, {Meyer},
  {Tucker}, {Wollack}, \& {Wright}}]{2008arXiv0811.4280D}
{Dunkley}, J., {Spergel}, D.~N., {Komatsu}, E., {et~al.} 2008{\natexlab{b}},
  ArXiv e-prints

\bibitem[{{Finkbeiner}(2004)}]{2004ApJ...614..186F}
{Finkbeiner}, D.~P. 2004, \apj, 614, 186

\bibitem[{{Finkbeiner} {et~al.}(1999){Finkbeiner}, {Davis}, \&
  {Schlegel}}]{1999ApJ...524..867F}
{Finkbeiner}, D.~P., {Davis}, M., \& {Schlegel}, D.~J. 1999, \apj, 524, 867

\bibitem[{{Fosalba} {et~al.}(2002){Fosalba}, {Lazarian}, {Prunet}, \&
  {Tauber}}]{2002ApJ...564..762F}
{Fosalba}, P., {Lazarian}, A., {Prunet}, S., \& {Tauber}, J.~A. 2002, \apj,
  564, 762

\bibitem[{{Giardino} {et~al.}(2002){Giardino}, {Banday}, {G{\'o}rski},
  {Bennett}, {Jonas}, \& {Tauber}}]{2002A&A...387...82G}
{Giardino}, G., {Banday}, A.~J., {G{\'o}rski}, K.~M., {et~al.} 2002, \aap, 387,
  82

\bibitem[{{Gold} {et~al.}(2008){Gold}, {Bennett}, {Hill}, {Hinshaw}, {Odegard},
  {Page}, {Spergel}, {Weiland}, {Dunkley}, {Halpern}, {Jarosik}, {Kogut},
  {Komatsu}, {Larson}, {Meyer}, {Nolta}, {Wollack}, \&
  {Wright}}]{2008arXiv0803.0715G}
{Gold}, B., {Bennett}, C.~L., {Hill}, R.~S., {et~al.} 2008, ArXiv e-prints, 803

\bibitem[{{Haslam} {et~al.}(1982){Haslam}, {Salter}, {Stoffel}, \&
  {Wilson}}]{1982A&AS...47....1H}
{Haslam}, C.~G.~T., {Salter}, C.~J., {Stoffel}, H., \& {Wilson}, W.~E. 1982,
  \aaps, 47, 1

\bibitem[{{Hirata} \& {Seljak}(2003)}]{2003PhRvD..67d3001H}
{Hirata}, C.~M. \& {Seljak}, U. 2003, \prd, 67, 043001

\bibitem[{{Hivon} {et~al.}(2002){Hivon}, {G{\'o}rski}, {Netterfield}, {Crill},
  {Prunet}, \& {Hansen}}]{2002ApJ...567....2H}
{Hivon}, E., {G{\'o}rski}, K.~M., {Netterfield}, C.~B., {et~al.} 2002, \apj,
  567, 2

\bibitem[{{Hu} {et~al.}(2003){Hu}, {Hedman}, \&
  {Zaldarriaga}}]{2003PhRvD..67d3004H}
{Hu}, W., {Hedman}, M.~M., \& {Zaldarriaga}, M. 2003, \prd, 67, 043004

\bibitem[{{Hu} \& {White}(1997)}]{1997NewA....2..323H}
{Hu}, W. \& {White}, M. 1997, New Astronomy, 2, 323

\bibitem[{{Jonas} {et~al.}(1998){Jonas}, {Baart}, \&
  {Nicolson}}]{1998MNRAS.297..977J}
{Jonas}, J.~L., {Baart}, E.~E., \& {Nicolson}, G.~D. 1998, \mnras, 297, 977

\bibitem[{{Kamionkowski} \& {Kosowsky}(1998)}]{1998PhRvD..57..685K}
{Kamionkowski}, M. \& {Kosowsky}, A. 1998, \prd, 57, 685

\bibitem[{{Kamionkowski} {et~al.}(1997){Kamionkowski}, {Kosowsky}, \&
  {Stebbins}}]{1997PhRvL..78.2058K}
{Kamionkowski}, M., {Kosowsky}, A., \& {Stebbins}, A. 1997, Physical Review
  Letters, 78, 2058

\bibitem[{{Kaplan} \& {Delabrouille}(2002)}]{2002AIPC..609..209K}
{Kaplan}, J. \& {Delabrouille}, J. 2002, in American Institute of Physics
  Conference Series, Vol. 609, Astrophysical Polarized Backgrounds, ed.
  S.~{Cecchini}, S.~{Cortiglioni}, R.~{Sault}, \& C.~{Sbarra}, 209--214

\bibitem[{{Kesden} {et~al.}(2003){Kesden}, {Cooray}, \&
  {Kamionkowski}}]{2003PhRvD..67l3507K}
{Kesden}, M., {Cooray}, A., \& {Kamionkowski}, M. 2003, \prd, 67, 123507

\bibitem[{{Knox} \& {Song}(2002)}]{2002PhRvL..89a1303K}
{Knox}, L. \& {Song}, Y.-S. 2002, Physical Review Letters, 89, 011303

\bibitem[{{Kovac} {et~al.}(2002){Kovac}, {Leitch}, {Pryke}, {Carlstrom},
  {Halverson}, \& {Holzapfel}}]{2002Natur.420..772K}
{Kovac}, J.~M., {Leitch}, E.~M., {Pryke}, C., {et~al.} 2002, \nat, 420, 772

\bibitem[{{Lazarian}(2007)}]{2007JQSRT.106..225L}
{Lazarian}, A. 2007, Journal of Quantitative Spectroscopy and Radiative
  Transfer, 106, 225

\bibitem[{{Lazarian} \& {Finkbeiner}(2003)}]{2003NewAR..47.1107L}
{Lazarian}, A. \& {Finkbeiner}, D. 2003, New Astronomy Review, 47, 1107

\bibitem[{{Lewis} \& {Challinor}(2006)}]{2006PhR...429....1L}
{Lewis}, A. \& {Challinor}, A. 2006, \physrep, 429, 1

\bibitem[{{Li} \& {Draine}(2001)}]{2001ApJ...554..778L}
{Li}, A. \& {Draine}, B.~T. 2001, \apj, 554, 778

\bibitem[{{Macellari} {et~al.}(2008){Macellari}, {Pierpaoli}, {Dickinson}, \&
  {Vaillancourt}}]{Macellari08}
{Macellari}, N., {Pierpaoli}, E., {Dickinson}, C., \& {Vaillancourt}, J. 2008,
  in preparation

\bibitem[{{Miville-Desch\^enes} {et~al.}(2008){Miville-Desch\^enes}, {Ysard},
  {Lavabre}, {Ponthieu}, {Macias-Perez}, {Aumont}, \&
  {Bernard}}]{2008arXiv0802.3345M}
{Miville-Desch\^enes}, M.~., {Ysard}, N., {Lavabre}, A., {et~al.} 2008, ArXiv
  e-prints, 802

\bibitem[{{Moudden} {et~al.}(2004){Moudden}, {Cardoso}, {Starck}, \&
  {Delabrouille}}]{2004astro.ph..7053M}
{Moudden}, Y., {Cardoso}, J.~., {Starck}, J.~., \& {Delabrouille}, J. 2004,
  ArXiv Astrophysics e-prints

\bibitem[{{North} {et~al.}(2008){North}, {Johnson}, {Ade}, {Audley}, {Baines},
  {Battye}, {Brown}, {Cabella}, {Calisse}, {Challinor}, {Duncan}, {Ferreira},
  {Gear}, {Glowacka}, {Goldie}, {Grimes}, {Halpern}, {Haynes}, {Hilton},
  {Irwin}, {Jones}, {Lasenby}, {Leahy}, {Leech}, {Maffei}, {Mauskopf},
  {Melhuish}, {O'Dea}, {Parsley}, {Piccirillo}, {Pisano}, {Reintsema},
  {Savini}, {Sudiwala}, {Sutton}, {Taylor}, {Teleberg}, {Titterington},
  {Tsaneva}, {Tucker}, {Watson}, {Withington}, {Yassin}, \&
  {Zhang}}]{2008arXiv0805.3690N}
{North}, C.~E., {Johnson}, B.~R., {Ade}, P.~A.~R., {et~al.} 2008, ArXiv
  e-prints

\bibitem[{{Page} {et~al.}(2007){Page}, {Hinshaw}, {Komatsu}, {Nolta},
  {Spergel}, {Bennett}, {Barnes}, {Bean}, {Dor{\'e}}, {Dunkley}, {Halpern},
  {Hill}, {Jarosik}, {Kogut}, {Limon}, {Meyer}, {Odegard}, {Peiris}, {Tucker},
  {Verde}, {Weiland}, {Wollack}, \& {Wright}}]{2007ApJS..170..335P}
{Page}, L., {Hinshaw}, G., {Komatsu}, E., {et~al.} 2007, \apjs, 170, 335

\bibitem[{{Pierpaoli} \& {Perna}(2004)}]{Pierpaoli04}
{Pierpaoli}, E. \& {Perna}, R. 2004, \mnras, 354, 1005

\bibitem[{{Platania} {et~al.}(2003){Platania}, {Burigana}, {Maino}, {Caserini},
  {Bersanelli}, {Cappellini}, \& {Mennella}}]{2003A&A...410..847P}
{Platania}, P., {Burigana}, C., {Maino}, D., {et~al.} 2003, \aap, 410, 847

\bibitem[{{Reich} \& {Reich}(1986)}]{1986A&AS...63..205R}
{Reich}, P. \& {Reich}, W. 1986, \aaps, 63, 205

\bibitem[{{Ricci} {et~al.}(2004){Ricci}, {Prandoni}, {Gruppioni}, {Sault}, \&
  {De Zotti}}]{2004A&A...415..549R}
{Ricci}, R., {Prandoni}, I., {Gruppioni}, C., {Sault}, R.~J., \& {De Zotti}, G.
  2004, \aap, 415, 549

\bibitem[{{Rosset} {et~al.}(2007){Rosset}, {Yurchenko}, {Delabrouille},
  {Kaplan}, {Giraud-H{\'e}raud}, {Lamarre}, \& {Murphy}}]{2007A&A...464..405R}
{Rosset}, C., {Yurchenko}, V.~B., {Delabrouille}, J., {et~al.} 2007, \aap, 464,
  405

\bibitem[{{Rybicki} \& {Lightman}(1979)}]{1979rpa..book.....R}
{Rybicki}, G.~B. \& {Lightman}, A.~P. 1979, {Radiative processes in
  astrophysics} (New York, Wiley-Interscience, 1979.~393 p.)

\bibitem[{{Sazonov} \& {Sunyaev}(1999)}]{1999MNRAS.310..765S}
{Sazonov}, S.~Y. \& {Sunyaev}, R.~A. 1999, \mnras, 310, 765

\bibitem[{{Schlegel} {et~al.}(1998){Schlegel}, {Finkbeiner}, \&
  {Davis}}]{1998ApJ...500..525S}
{Schlegel}, D.~J., {Finkbeiner}, D.~P., \& {Davis}, M. 1998, \apj, 500, 525

\bibitem[{{Seljak} \& {Zaldarriaga}(1997)}]{1997PhRvL..78.2054S}
{Seljak}, U. \& {Zaldarriaga}, M. 1997, Physical Review Letters, 78, 2054

\bibitem[{{Seto} \& {Pierpaoli}(2005)}]{Seto05}
{Seto}, N. \& {Pierpaoli}, E. 2005, Physical Review Letters, 95, 101302

\bibitem[{{Sievers} \& {CBI Collaboration}(2005)}]{2005AAS...20710007S}
{Sievers}, J. \& {CBI Collaboration}. 2005, in Bulletin of the American
  Astronomical Society, Vol.~37, Bulletin of the American Astronomical Society,
  1329--+

\bibitem[{{Smith} {et~al.}(2008){Smith}, {Cooray}, {Das}, {Dor{\'e}}, {Hanson},
  {Hirata}, {Kaplinghat}, {Keating}, {LoVerde}, {Miller}, {Rocha}, {Shimon}, \&
  {Zahn}}]{2008arXiv0811.3916S}
{Smith}, K.~M., {Cooray}, A., {Das}, S., {et~al.} 2008, ArXiv e-prints

\bibitem[{{Strong} {et~al.}(2007){Strong}, {Moskalenko}, \&
  {Ptuskin}}]{2007ARNPS..57..285S}
{Strong}, A.~W., {Moskalenko}, I.~V., \& {Ptuskin}, V.~S. 2007, Annual Review
  of Nuclear and Particle Science, 57, 285

\bibitem[{{Tegmark} {et~al.}(2003){Tegmark}, {de Oliveira-Costa}, \&
  {Hamilton}}]{2003PhRvD..68l3523T}
{Tegmark}, M., {de Oliveira-Costa}, A., \& {Hamilton}, A.~J. 2003, \prd, 68,
  123523

\bibitem[{{Tucci} {et~al.}(2004){Tucci}, {Mart{\'{\i}}nez-Gonz{\'a}lez},
  {Toffolatti}, {Gonz{\'a}lez-Nuevo}, \& {De Zotti}}]{tucci/etal:2004}
{Tucci}, M., {Mart{\'{\i}}nez-Gonz{\'a}lez}, E., {Toffolatti}, L.,
  {Gonz{\'a}lez-Nuevo}, J., \& {De Zotti}, G. 2004, \mnras, 349, 1267

\bibitem[{{Tucci} {et~al.}(2005){Tucci}, {Mart{\'{\i}}nez-Gonz{\'a}lez},
  {Vielva}, \& {Delabrouille}}]{2005MNRAS.360..935T}
{Tucci}, M., {Mart{\'{\i}}nez-Gonz{\'a}lez}, E., {Vielva}, P., \&
  {Delabrouille}, J. 2005, \mnras, 360, 935

\bibitem[{{Verde} {et~al.}(2006){Verde}, {Peiris}, \&
  {Jimenez}}]{2006JCAP...01..019V}
{Verde}, L., {Peiris}, H.~V., \& {Jimenez}, R. 2006, Journal of Cosmology and
  Astro-Particle Physics, 1, 19

\bibitem[{{Zaldarriaga} \& {Seljak}(2000)}]{2000ApJS..129..431Z}
{Zaldarriaga}, M. \& {Seljak}, U. 2000, \apjs, 129, 431

\bibitem[{{Zaldarriaga} {et~al.}(1998){Zaldarriaga}, {Seljak}, \&
  {Bertschinger}}]{1998ApJ...494..491Z}
{Zaldarriaga}, M., {Seljak}, U., \& {Bertschinger}, E. 1998, \apj, 494, 491

\end{thebibliography}

\appendix{}

\section{Parameterisation the foreground component and choice of a mask}
\label{sec:d}

In this appendix, we discuss in more detail the dimension $D$ of
matrix used to represent the covariance of the total galactic
emission, and the choice of a mask to hide regions of strong galactic
emission for the estimation of $r$ with \smica.

\subsection{Dimension $D$ of the foreground component}

First, we explain on a few examples the mechanisms which set the rank
of the foreground covariance matrix, to give an intuitive
understanding of how the dimension $D$ of the foregrounds component
used in \smica\ to obtain a good model of the data.

Let's consider the case of a `perfectly coherent' physical process, for
which the total emission, as a function of sky direction $\xi$ and
frequency $\nu$, is well described by a spatial template multiplied by
a pixel-independent power law frequency scaling:
\begin{equation}
S_\nu(\xi) = S_0(\xi) \left( \frac{\nu}{\nu_0} \right)^{\beta}
\label{eq:monodim}
\end{equation}

The covariance matrix of this foreground will be of rank one and
\(\R[S] = [ \mathbf{A} \mathbf{A}^\dag \var (S_0)] \), with \(A_f =
\left( \frac{\nu_f}{\nu_0} \right)^{\beta} \). Now, if the spectral
index \(\beta\) fluctuates on the sky, \(\beta(\xi) = \beta +
\delta\beta(\xi)\), to first order, the emission at frequency \(\nu\)
around \(\nu_0\) can be written:
\begin{equation}
S_\nu(\xi) \approx S_0(\xi) \left( \frac{\nu}{\nu_0} \right)^{\beta} + S_0(\xi) \left( \frac{\nu}{\nu_0} \right)^{\beta} \delta \beta(\xi) \left( \frac{\nu - \nu_0}{\nu_0}\right)
\label{eq:bidim}
\end{equation}

This is not necessarily the best linear approximation of the emission,
but supposing it holds, the covariance matrix of the foreground will
be of rank two (as the sum of two correlated rank 1 processes). If the
noise level is sufficiently low, the variation introduced by the first
order term of Eq. \ref{eq:bidim} becomes truly significant, we can't
model the emission by a mono-dimensional component as in
Eq.\ref{eq:monodim}.

In this work, we consider two processes, synchrotron and dust, which
are expected to be correlated (at least by the galactic magnetic field
and the general shape of the galaxy). Moreover, significant spatial
variation of their emission law arises (due to cosmic aging, dust
temperature variation ...), which makes their emission only partially
coherent from one channel to another. Consequently, we expect that the
required dimension \(D\) of the galactic foreground component will be
at least 4 as soon as the noise level of the instrument is low enough.

The selection of the model can also be made on the basis of a
statistical criterion. For example, Table \ref{tab:bic} shows the
Bayesian information criterion (BIC) in the case of the EPIC-2m
experiment ($r = 0.01$) for 3 consecutive values of $D$. The BIC is a
decreasing function of the likelihood and of the number of
parameter. Hence, lower BIC implies either fewer explanatory
variables, better fit, or both. In our case the criterion reads:
\[
BIC = -2 \loglik + k \ln\sum_q w_q
\]
where $k$ is the number of estimated parameters and $w_q$ the
effective number of modes in bin $q$. Taking into account the
redundancy in the parameterisation, the actual number of free
parameters in the model is $ 1 + F\times D + Q{D(D+1)}/ 2 - D^2$.
However, we usually prefer to rely on the inspection of the mismatch
in every bin of $\ell$, as some frequency specific features may be
diluted in the global mismatch.

\begin{table}
  \centering\[
  \begin{array}{lccc}
    \hline
    \hline
    D & k & BIC \\
    \hline
     3 & 376 & 1.15 \cdot 10^{4} \\ 4 & 617 & 8.35 \cdot 10^{3} \\ 5 & 916 & 1.15 \cdot 10^{4} \\
    \hline
  \end{array}\]
  \caption{Bayesian information criterion of 3 models with increasing dimension of the galactic component for the EPIC-2m mission. The selected value $D = 4$ correspond to a minimum of this criterion.}
  \label{tab:bic} 
\end{table}

\subsection{Masking influence}
\label{sec:maskstud}

The noise level and the scanning strategy remaining fixed in the
full-sky experiments, a larger coverage gives more information and
should result in tighter constraints on both foreground and CMB. In
practice, it is only the case up to a certain point, due to the non
stationarity of the foreground emission. In the galactic plane, the
emission is too strong and too complex to fit in the proposed model,
and this region must be discarded to avoid contamination of the
results.
The main points governing the choice of an appropriate mask are the
following:

\begin{itemize} 
\item The covariance of the total galactic emission (synchrotron and
  dust polarised emissions), because of the variation of emission laws
  as a function of the direction on the sky, is never \emph{exactly}
  modelled by a rank $D$ matrix. However it is 
  \emph{satisfactorily} modelled in this way if the difference between
  the actual second order statistics of the foregrounds, and those of
  the rank $D$ matrix model, are indistinguishable because of the
  noise level (or because of cosmic variance in the empirical
  statistics).  The deviation from the model is more obvious in
  regions of strong galactic emission, hence the need for a galactic
  mask. The higher the noise, the smaller the required mask.
\item \smica\ provides a built-in measure of the adequacy of the model, which is the value of the spectral mismatch. If too high, the model under-fits the data, and the dimension of the foreground model (or the size of the mask) should be increased. If too low, the model over-fits the data, and $D$ should be decreased.
\item Near full sky coverage is better for measuring adequately the reionisation bump.
\item The dimension of the foreground component must be smaller than
  the number of channels.
\end{itemize}


If the error variance is always dominated by noise and cosmic
variance, the issue is solved: one should select the smaller mask that
gives a good fit between the model and the data to minimise the mean
squared error and keep the estimator unbiased.

If, on the other hand, the error seems dominated by the contribution
of foregrounds, which is, for example, the case of the EPIC-2m
experiment for $r = 0.001$, the tradeoff is unclear and it may happen
that a better estimator is obtained with a stronger masking of the
foreground contamination. We found that it is not the case. Table
\ref{tab:mask} illustrates the case of the EPIC-2m experiment with the
galactic cut used in Sect. \ref{sec:results} and a bigger
cut. Although the reduction of sensitivity is slower in presence of
foreground than for the noise dominated case, the smaller mask still
give the better results.

\begin{table}
  \[
  \begin{array}{c|cccc}
    \hline
    \hline
    r & r^\text{est} & \sigma_r^{\rm FISHER} & \sigma_r^{\rm no-fg} & \fsky \\
    \hline
     0.001 & 1.01 \cdot 10^{-3} & 1.60 \cdot 10^{-4} & 5.25 \cdot 10^{-5} & 0.87\\
     0.001 & 1.01 \cdot 10^{-3} & 1.68 \cdot 10^{-4} & 5.72 \cdot 10^{-5} & 0.73\\
    \hline
  \end{array}
  \]
  \caption{Estimation of the tensor to scalar ratio with two different galactic cuts in the EPIC-2m experiment.}
  \label{tab:mask}
\end{table}
We may also recall that the expression~(\ref{eq:loglikesingle}) of the
likelihood is an approximation for partial sky coverage. The scheme
presented here thus may not give fully reliable results when masking
effects become important.



\section{Spectral mismatch}
\label{sec:mm}

\begin{figure}
  \includegraphics{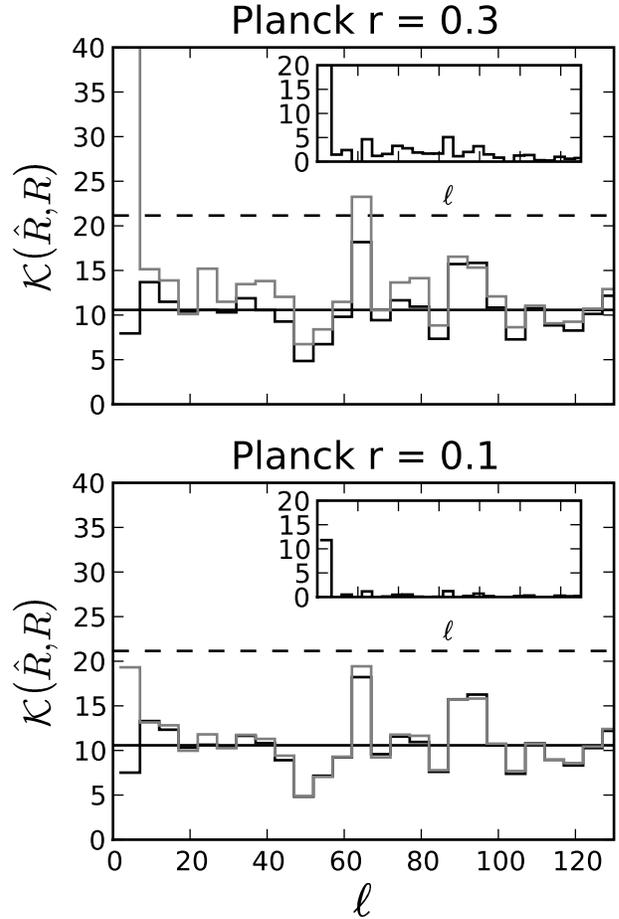}
  \caption{Those plots present the distribution in \(\ell\) of the
    mismatch criterion between the model and the data for two values
    of \(r\) for \textsc{Planck}. On the grey curve, the mismatch has
    been computed discarding the CMB contribution from the \smica\
    model. The difference between the two curves, plotted in
    inclusion, illustrates somehow the importance of the CMB
    contribution to the signal.}
  \label{fig:planckmismatch}
\end{figure}

\begin{figure*}
  \includegraphics{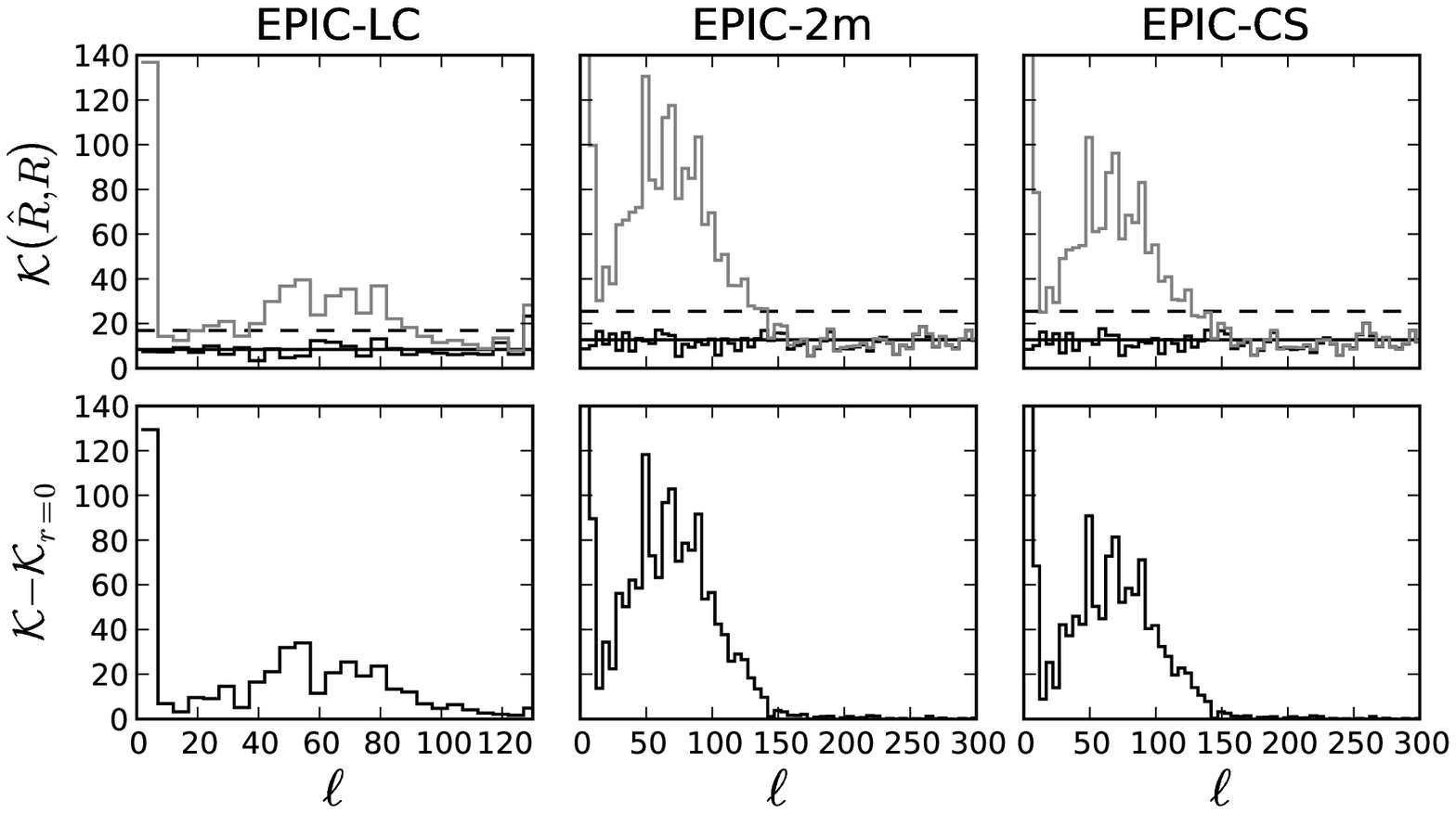}
  \includegraphics{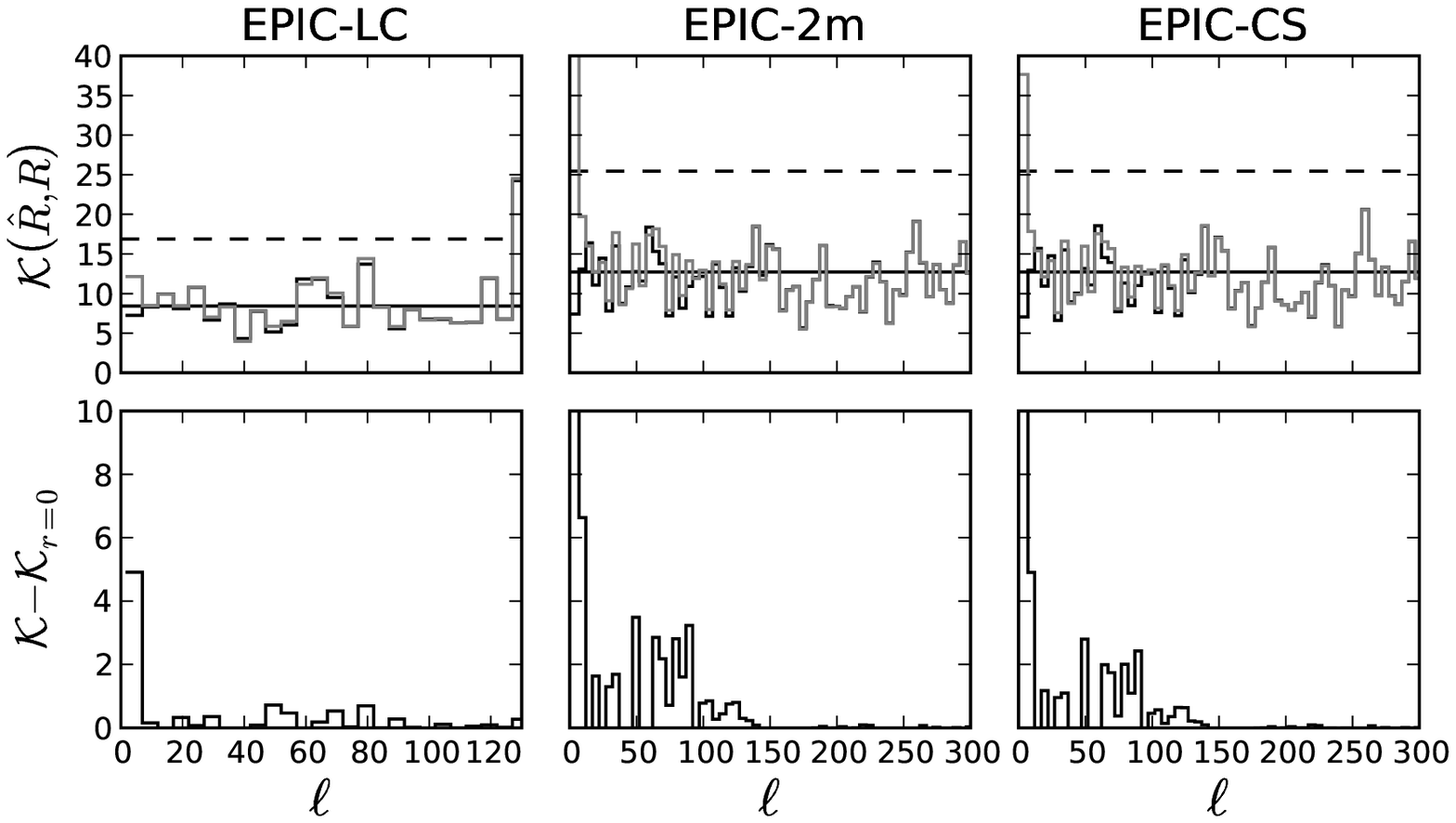}
  \caption{Mismatch criterion for \(r = 0.01\) (top) and \(r = 0.001\) (bottom). In each plot, the top panel shows the mismatch criterion between the best fit model and the data (black curve) and the best fit model deprived from the CMB contribution and the data (gray curve). Solid and dashed horizontal lines show respectively the mismatch expectation and 2 times the mismatch expectation. The difference between the gray and the black curve is plotted in the bottom panel and gives an idea of the significance of the CMB signal in each bin of $\ell$.}
  \label{fig:epicmm}
\end{figure*}

Computed for each bin $q$ of $\ell$, the mismatch criterion, $w_q
\KL{\R*_q}{\R_q(\theta^*)}$, between the best-fit model
$\R_q(\theta^*)$ at the point of convergence $\theta^*$, and the data
$\R*_q$, gives a picture of the goodness of fit as a function of the
scale. Black curves in Figs. \ref{fig:planckmismatch} and
\ref{fig:epicmm} show the mismatch criterion of the best fits for
Planck and EPIC designs respectively. When the model holds, the value
of the mismatch is expected to be around the number of degrees of
freedom (horizontal black lines in the figures). We can also compute
the mismatch for a model in which we discard the CMB contribution $w_q
\KL{\R*_q}{\R_q(\theta^*) - \R[CMB]_q(r^*)}$. Gray curves in
Figs. \ref{fig:planckmismatch} and \ref{fig:epicmm} show the mismatch
for this modified model. The difference between the two curves
illustrates the `weight' of the CMB component in the fit, as a
function of scale.

Figure \ref{fig:planckmismatch} shows the results for Planck for
$r=0.3$ and 0.1. The curves of the difference plotted in inclusion
illustrate the predominance of the reionisation bump. In
Fig. \ref{fig:epicmm}, we plot the difference curve on the bottom
panels for the three experiments for $r=0.01$ and $r=0.001$. They
illustrate clearly the difference of sensitivity to the peak between
the EPIC-LC design and the higher resolution experiments. In general
it can be seen that no significant contribution to the CMB is coming
from scales smaller than $\ell = 150$.

\end{document}